\documentclass[review]{elsarticle}
\usepackage{lipsum,caption}
\usepackage{algorithmic}
\makeatletter
\def\ps@pprintTitle{
\let\@oddhead\@empty
\let\@evenhead\@empty
\def\@oddfoot{}%
\let\@evenfoot\@oddfoot}
\makeatother
\usepackage{mathtools}

\usepackage[utf8]{inputenc}
\usepackage[margin=2.5cm]{geometry}
\usepackage{multirow}
\usepackage{float}
\usepackage{lipsum}
\usepackage{amsfonts}
\usepackage{amsthm}
\usepackage{amsmath,scalerel}
\usepackage{longtable}
\usepackage{caption}
\usepackage{booktabs}
\usepackage{rotating}
\usepackage{pdflscape}
\usepackage{bm}
\usepackage{epstopdf}
\usepackage{titlesec} 
\usepackage{float} 

\usepackage{algorithm2e}
\usepackage{fixmath}
\usepackage{amssymb}\usepackage{graphicx}
\usepackage[english]{babel}
\usepackage{mathtools}
\usepackage{tikz, wrapfig,array}
\usetikzlibrary{calc,patterns,
	decorations.pathmorphing,
	decorations.markings, shapes.geometric, graphs, graphs.standard, quotes,shapes,chains,scopes,positioning,arrows}
	
\usetikzlibrary{arrows.meta}
 
\usepackage[tight,footnotesize]{subfigure}
\usepackage{enumitem}
\usepackage{arydshln}

\usetikzlibrary{trees}
\newtheorem{theorem}{Theorem}[section]

\newtheorem{lemma}[]{Lemma}

\theoremstyle{definition}

\newtheorem{remark}[]{Remark}
\newtheorem{remark.ap}[]{Remark}
\theoremstyle{definition}

\newcommand{\GG}[1]{}
\usetikzlibrary{arrows}

\begin{document}

\title{On the use of historical estimates}
\author[]{Ori Davidov}
\address{Department of Statistics, University of Haifa, Mount Carmel, Haifa 3498838, Israel}
\author{Tam\'{a}s Rudas} 
\address{Department of Statistics, Faculty of Social Sciences, E\"{o}tv\"{o}s Lor\'{a}nd University}  
\address{P\'azm\'any P\'eter s\'et\'any 1/A, Budapest, Hungary}

\begin{abstract}
The use of historical estimates in current studies is common in a wide variety of application areas. Nevertheless, despite their routine use the uncertainty associated with historical estimates is rarely properly accounted for in the analysis. In this communication we review common practices and then provide a mathematical formulation and a principled methodology for addressing the problem of drawing inferences in the presence of historical data. Three distinct variants are investigated in detail; the corresponding limiting distributions are found and compared. The design of future studies, given historical data, is also explored and relations with a variety of other well--studied statistical problems discussed. 
\end{abstract}

\bigskip
	
	
\maketitle
\date{\today}
\section{Introduction}\label{Sec.Intro}

There are many circumstances in which a statistical analysis either requires, or can greatly benefit, from the use of historical data. Often, the historical data consists of parameter estimates, which are essential for model fitting but impossible, or very expensive, to collect in the context of the current study. As reviewed below historical data are used in a variety of applications in the social, physical, and biomedical sciences. 

The planning of early detection programs for breast, prostate or other cancers requires knowledge of the sojourn time distributions in the healthy, pre--clinical, and clinical states. The sojourn times, however, are not directly observed, rather their convolution, i.e, the overall incidence rate is observed. Cancer incidence rates are collected by various cancer agencies and registries such as the Surveillance, Epidemiology and End Results (SEER) registry maintained by the US National Cancer Institute (https://seer.cancer.gov/registries/). Given the overall incidence rate and the estimated values of some parameters all sojourn time distributions can be then estimated by deconvolution. See Lee and Zelen (1998) and Davidov and Zelen (2004) for further details. It is worth noting that the application above may be viewed as a model for situations in which knowledge collected in one setting, experimental or observational, is then used to estimate quantities arising in a different experiment and is quite common in the biomedical sciences. For example, some models for the spread of infectious diseases, such as the SIR model (Becker, 2017) require the input of age specific transmission parameters which can be estimated from social contact networks (Edmunds et al. 1997, Wallinga et al. 2006) and then used to fit epidemic models (Mossong et al. 2008, Goeyvaerts et al. 2010, Yaari et al. 2016). Another interesting application is the optimization of cancer treatment where Kronick et al. (2010) develop a framework for predicting the outcome of prostate cancer immunotherapy by fitting personalized mathematical models. Their model consists of a set of differential equations whose behavior is governed by a collection of parameters, some of which are global parameters while others are subject specific. The values of the global parameters were obtained from at least ten different published studies, see their Table 2, whereas the subject level parameters were estimated by fitting a model to each participant assuming that the global parameters were estimated without error. See Kogan et al. (2012) and Kozłowska et al. (2018) for a similar applications. Another very important application in which historical data is used is clinical trials. Consider, for example, the situation in which the effect of a combination of treatments is assessed (e.g., Tamma et al. 2012, Kandra et al. 2016). In such cases there exists a collection of therapies which have been independently proven to be somewhat successful at treating a medical condition. The objective of a new study may then be to assess whether a combination of these therapies provides an even better outcome. In the simplest case, one may view this problem as a three armed clinical trial comparing treatments $\mathbold{A}$,$\mathbold{B}$ and $\mathbold{A+B}$ in which historical data on treatments $\mathbold{A}$ and $\mathbold{B}$ already exists. An important example of such situations is the Federal Drug Administration (2006) guidelines for submitting applications for approval of fixed dose combinations, i.e., co--packaged drug products, of previously approved antiretrovirals for the treatment of HIV. In particular, Attachment A of the aforementioned document considers the scenario in which a non--innovator, i.e., a generic drug company, wants to obtain approval for a combination of already approved ingredients. In this case, only efficacy data for the combination needs to be submitted. We will revisit and thoroughly analyze two forms of this example later on. More broadly, the use of historical data in the contexts of clinical trials has been investigated by numerous researchers and multiple perspectives, cf., Pocock (1976), Peto et al. (1979), Neuenschwander et al. (2010), Viele et al. (2014), and Piantadosi (2017) among many others. 

The use of historical estimates is also widespread in the social sciences. For example, in the fitting of some econometric models researchers may use values estimated from previously collected survey data. The complexity of using historical estimates in the social sciences is well illustrated by the work of Tasseva (2019). In a microsimulation study investigating the effect of the recent expansion in higher education in Great Britain on household inequalities, previously obtained estimates of population figures from the Family Resources Survey for Great Britain (GOV.UK 2019) were used. The study used bootstrap methods in an effort to account for sampling variability in the survey data. But as noted by the author this method does not appropriately account for measurement error, inevitably present in income information collected in surveys, see, e.g., Moore et al. (2000). Similarly, Douidich et al. (2015) describe an imputation--related--method for incorporating estimates obtained in labor force surveys (which are easily and cheaply conducted) into household expenditure surveys (which are much more time consuming and expensive) in order to estimate poverty rates in Morocco. Likewise, demographic model fitting and projections rely on historical data. The standard method of population projections (see United Nations, 2014) is based on the combination of cohort survival rates, i.e., historical data, with current data on cohort sizes. Raftery et al. (2014) proposed a Bayesian approach to take the uncertainty associated with historical data into account. It is worth noting that in this case the uncertainty accounted for by the Bayesian modelling did not come from observational errors but rather from the fact, that the true population figures may have changed over time.

We observe that researchers often do not adequately account for the variability of the historical estimators when incorporating them into a current analysis. In fact, we believe that the practice of plugging--in the estimated values for certain parameters is widespread. Nevertheless it is poorly accounted for, nor properly reported on, in the literature as most practitioners view this strategy as a natural way of "doing things". Therefore, the objectives of this communication are twofold: first to draw attention to current practice, and secondly and more importantly, to provide a principled methodology for incorporating uncertainty into analyses using historical estimates. Given the structure of the statistical problem we investigate various ways of doing so. The resulting methods are compared in terms of the precision of any estimated function of the model parameters and a preference order among them is established.  

The paper is organized as follows. Our notation and formulation are outlined in Section \ref{Sec.Notation.Formulation}. Section \ref{Sec.Results} provides our main theoretical findings which include the limiting distributions of the estimates in the presence of historical data and a comparison thereof. In Section \ref{Sec.Application} two applications are described in conjunction with accompanying numerical experiments. The first application addresses the two--way analysis of variance (ANOVA) problem introduced in Section \ref{Sec.Notation.Formulation}. The second, related application, deals with a drug interaction study within the framework of Bliss--independence (Bliss, 1939), an old concept which has garnered much recent attention. We conclude with a discussion in Section \ref{Sec.Disscusion}. All proofs are collected in an Appendix.

\section{Notation and formulation} \label{Sec.Notation.Formulation}

Consider a designed experiment or observational study, denoted by $\mathcal{S}$, in which data $\mathcal{D}$ of size $n$ is observed. Suppose further that the model describing the distribution of $\mathcal{D}$ is indexed by $\bm{\omega}^T=(\bm{\theta}^T,\bm{\eta}^T)$ where $\bm{\theta}\in \mathbb{R}^{p}$ and $\bm{\eta}\in \mathbb{R}^{q}$ is the concatenation of    
$\bm{\eta}_{1}\in\mathbb{R}^{q_1},\ldots,\bm{\eta}_{K}\in\mathbb{R}^{q_K}$ with $q=q_1+\cdots+q_K$. Let $\bm{\Phi}(\bm{\omega})$ be some function of the model parameters which is of interest to the researchers. Clearly $\bm{\Phi}(\bm{\omega})$ may be a function of $\bm{\theta}$ alone, $\bm{\eta}$ alone or of both $\bm{\theta}$ and $\bm{\eta}$. The primary goal of the study $\mathcal{S}$, which we refer to as the current study, is inference on $\bm{\Phi}(\bm{\omega})$ in the presence of historical data which we view as a collection of $K$, independent estimators $\widehat{\bm{\eta}}_{1},\ldots,\widehat{\bm{\eta}}_{K}$ obtained from historical studies $\mathcal{S}_1,\ldots,\mathcal{S}_K$ of sizes $m_1,\ldots,m_K$ and $m= m_1+\cdots+m_K$ denotes the total sample size in the historical studies. 

In some circumstances the statistical model for $\mathcal{D}$ may not be sufficient to identify $\bm{\omega}$ but would allow the estimation of $\bm{\theta}$ if $\bm{\eta}$ were known in advance (e.g. Kronick et al., 2009). In other circumstances given the data $\mathcal{D}$ both $\bm{\theta}$ and $\bm{\eta}$ are estimable (e.g., Peddada et al., 2007). Thus, in this communication we consider two distinct settings, the second of which has two variants. In the first setting, referred to as a Type {\rm I} Problem, only the parameter $\bm{\theta}$ is estimated using the data $\mathcal{D}$, while $(\bm{\eta}_{1},\ldots,\bm{\eta}_{K})$ are fixed at their historical estimated values $(\widehat{\bm{\eta}}_{1},\ldots,\widehat{\bm{\eta}}_{K})$. In the second setting, referred to as Type {\rm II} Problem, a two--step procedure is utilized to estimate $\bm{\theta}$ while updating the estimators for $(\bm{\eta}_{1},\ldots,\bm{\eta}_{K})$. In some cases, although a Type {\rm II} analysis would be possible, the researcher may decide not to do so and apply a Type {\rm I} analysis. One of our results shows that this is an inferior strategy, i.e., if the data $\mathcal{D}$ identifies $\bm{\omega}$ it is always advisable to re--estimate $\bm{\eta}$. It is also important to emphasize that there are situations in which the investigator, by means of the design of the study $\mathcal{S}$, may control whether the problem is of Type {\rm I} or a Type {\rm II}. 

To fix ideas consider the two--way ANOVA model in which the expected value of an outcome $Y$ is given by
\begin{equation} \label{eq.twoway.anova}
\mathbb{E}(Y|T_{1},T_{2}) = \eta_{0} + \eta_{1}T_{1} + \eta_{2}T_{2} + \theta T_{1}T_{2} 
\end{equation}
where $T_{i}\in\{0,1\}$ for $i=1,2$ indicates whether treatment $i$ is administered. Here $\eta_{0}$ denotes the mean of $Y$ when neither treatment is administered, $\eta_{i}$ models the marginal increase in the expectation of $Y$ when treatment $i$ is administered and $\theta$ models the interaction $T_{1} \times T_{2}$. Suppose, now that the historical data consists of two studies $\mathcal{S}_1$ and $\mathcal{S}_2$ of sizes $m_1$ and $m_2$, respectively, where in the study $\mathcal{S}_i$ treatment $i$ was compared with a control. Clearly the historical data provides no information on $\theta$. Thus inference on $\theta$ would require a new study $\mathcal{S}$ in which $T_{1} = T_{2} = 1$ for some subset of the observations. For simplicity, {\it interchangeability} is assumed, i.e., all experimental units, in $\mathcal{S}_1$ and $\mathcal{S}_2$ as well as $\mathcal{S}$, are assumed to be drawn from the same population, e.g., Peddada et al. (2007), and therefore any change in the mean response may be attributed solely to the treatment combination received. The assumption of {\it interchangeability} may be relaxed as discussed in Section \ref{Sec.Disscusion}.

One objective of this communication is to provide a methodology for effective design and analysis of a new study $\mathcal{S}$ of size $n$ which allows the estimation of $\theta$ and utilizes the historical estimates of $(\eta_{0},\eta_{1},\eta_{2})$ obtained from $\mathcal{S}_1$ and $\mathcal{S}_2$. Depending on its objectives, the study $\mathcal{S}$ may be of various forms. For example, one may choose to allocate all $n$ observations to receive both treatments, i.e., $T_1=T_2=1$. In this case the data $\mathcal{D}$ is an IID sample of observations with mean $\eta_0+\eta_1+\eta_2+\theta$ and fixed variance $\sigma^2$. Although the parameter $\theta$ is not identifiable from $\mathcal{D}$ alone it is estimable given the historical data, so this is clearly a Type {\rm I} Problem.  Alternatively, if $\mathcal{S}$ allocates observation to all treatment combinations then $\theta$ as well as $(\eta_0,\eta_1,\eta_2)$ are estimable from $\mathcal{S}$ and this falls within the framework of a Type {\rm II} Problem. This example will be further analyzed in Section \ref{Sec.Application.ANOVA}. 

\section{Results} \label{Sec.Results}

Our main theoretical findings, i.e., Theorems \ref{Thm.Limit.4.Estimator.I}, \ref{Thm.Limit.4.Estimator.IIa}, and \ref{Thm.Limit.4.Estimator.IIb} describe the limiting distributions of estimators for $\bm{\omega}$ which are then compared in Theorems \ref{Thm.Order.A&B}, \ref{Thm.Order.C&U} and \ref{Thm.Order.B&C}.  

\subsection{Type {\rm I} Problems} \label{Sec.3.1}

Suppose first that we are in the setting of a Type {\rm I} Problem. Recall that in such circumstances only $\bm{\theta}$ is estimated while $(\bm{\eta}_1,\ldots,\bm{\eta}_K)$ are fixed at their historical values. Thus, let $\bar{\bm{\theta}}_{A}$ solve
\begin{equation}\label{Eq.Type.I}
\bm{\Psi}(\bm{\theta},\widehat{\bm{\eta}})=\bm{0}
\end{equation}
where $\widehat{\bm{\eta}}=(\widehat{\bm{\eta}}_{1}^{T},\ldots,\widehat{\bm{\eta}}_{K}^{T})^{T}$. The estimating equation \eqref{Eq.Type.I} may be a score equation motivated by likelihood theory, a generalized estimating equation derived by quasi--likelihood or any other statistical estimation framework. Observe that the solution $\bar{\bm{\theta}}_{A}$ of \eqref{Eq.Type.I} is obtained by plugging--in the sample values of the $K$ independent estimators $\widehat{\bm{\eta}}_{1},\ldots,\widehat{\bm{\eta}}_{K}$. For simplicity we may further assume that the data $\mathcal{D}$ is a random sample $\bm{Y}_{1},\ldots,\bm{Y}_{n}$ and \eqref{Eq.Type.I} is of the form
$$\bm{\Psi}(\bm{\theta},\widehat{\bm{\eta}})=n^{-1}\sum_{i=1}^{n}\bm{\psi}(\bm{\theta},\widehat{\bm{\eta}},\bm{Y}_{i}).$$
The function $\bm{\psi}$ is assumed to be: $(i)$ continuously differentiable with respect to both $\bm{\theta}$ and $\bm{\eta}_{1},\ldots,\bm{\eta}_{K}$; it is further assumed to satisfy $(ii)$ $\mathbb{E}_{0}(\bm{\psi}) = \bm{0}$; $(iii)$ $\mathbb{E}_{0}(\bm{\psi}\bm{\psi}^T) < \infty$; $(iv)$ the matrix $\mathbb{E}_{0}(\partial\bm{\psi}/\partial\bm{\eta})$ exist; and $(v)$ the matrix $\mathbb{E}_{0}(\partial\bm{\psi}/\partial\bm{\theta})$ exists and is invertible. Here $\mathbb{E}_{0}(\cdot)$ denotes the expectation taken at $\bm{\omega}_{0}=(\bm{\theta}_0,\bm{\eta}_{0})=(\bm{\theta}_0,\bm{\eta}_{1,0},\ldots,\bm{\eta}_{K,0})$, the true value of all parameters. Conditions $(i)-(v)$ are all standard regularity conditions often imposed in the literature (cf., Heyde 1997, Van der Vaart 2000). We now have the following: 

\begin{theorem} \label{Thm.Limit.4.Estimator.I}
Let $\bar{\bm{\theta}}_{A}$ be a solution to \eqref{Eq.Type.I} and set 
$\bar{\bm{\eta}}_{A}=\widehat{\bm{\eta}}$. Assume that: (i) $\bar{\bm{\theta}}_{A}$ is consistent at $\bm{\omega}_{0}$; (ii) the estimating function $\bm{\psi}$ satisfies the regularity conditions listed above; and (iii) the historical estimates satisfy $\sqrt{m_j}(\widehat{\bm{\eta}}_{j}-\bm{\eta}_{j,0}) \Rightarrow \mathcal{N}_{q_{j}}(\bm{0},\bm{\Sigma}_{j})$ and are independent of each other and of the current study. Then if $(m/m_j)\rightarrow\kappa_j < \infty$ for all $j=1,\ldots,K$ as  $m_j \to \infty$ and $n/m \rightarrow \rho \in (0,\infty)$ as $n \to \infty$ we have
\begin{equation*}
\sqrt{n}(\bar{\bm{\theta}}_{A}-\bm{\theta}_{0},\bar{\bm{\eta}}_{A}-\bm{\eta}_{0})^T \Rightarrow \mathcal{N}_{p+q}(\bm{0},\bm{A})
\end{equation*}
where 
\begin{equation*} \label{Eq.A}
\bm{A} =   
\begin{pmatrix}
\bm{A}_{\bm{\theta\theta}}  & \bm{A}_{\bm{\theta\eta}} \\
\bm{A}_{\bm{\eta\theta}}  & \bm{A}_{\bm{\eta\eta}}
\end{pmatrix}
\end{equation*}
with
\begin{align*}
\bm{A}_{\bm{\theta\theta}}&=(\bm{D}^{-1}_{\bm{\theta}_0})[\bm{\Sigma}_{\bm{\psi}} + \rho \bm{D}_{\bm{\eta}_0} \bm{\Sigma} \bm{D}^{T}_{\bm{\eta}_0}] (\bm{D}^{-1}_{\bm{\theta}_0})^{T},\\
\bm{A}_{\bm{\theta\eta}}&=-\rho(\bm{D}_{\bm{\theta}_0})^{-1}\bm{D}_{\bm{\eta}_0}\bm{\Sigma},\\
\bm{A}_{\bm{\eta\eta}}&=\rho\bm{\Sigma}.
\end{align*}
where 
$\bm{D}_{{\bm{\theta}_0}}=\mathbb{E}_{0}(\partial\bm{\psi}/\partial{\bm{\theta})}$, $\bm{D}_{{\bm{\eta}_0}}=\mathbb{E}_{0}(\partial\bm{\psi}/\partial{\bm{\eta })}$, $\bm{\Sigma}_{\bm{\psi}}=\mathbb{E}_{0}(\bm{\psi}\bm{\psi}^T)$ and $\bm{\Sigma} = \mathrm{BlockDiag}(\kappa _{1}\bm{\Sigma }_{1},\ldots,\kappa _{K}\bm{\Sigma}_{K})$.
\end{theorem}

\begin{remark}
Clearly, $\bm{A}_{\bm{\theta\theta}}$ is the $p\times p$ asymptotic variance matrix of $\bar{\bm{\theta}}_{A}$, $\bm{A}_{\bm{\eta\eta}}$ is the $q\times q$ asymptotic variance matrix of $\bar{\bm{\eta}}_{A}$ and $\bm{A}_{\bm{\theta\eta}} = \bm{A}_{\bm{\eta\theta}}^T$ is their $p\times q$ asymptotic covariance matrix.
\end{remark}

The proof of Theorem \ref{Thm.Limit.4.Estimator.I} is a straightforward, but somewhat involved, application of the delta method. In contrast with Randles (1982) and Pierce (1982) which describe the limiting distribution of statistics that are explicit functions of estimated parameters the estimator $\bar{\bm{\theta}}_{A}$ is an implicit function of $\bar{\bm{\eta}}_{A}$. Further note that $(\bm{D}^{-1}_{{\bm{\theta}_0}})\bm{\Sigma}_{\bm{\psi}}(\bm{D}^{-1}_{{\bm{\theta}_0}})^{T}$ is the asymptotic variance of $\bar{\bm{\theta}}_{A}$ when the true values of $\bm{\eta}_{1},\ldots,\bm{\eta}_{K}$ are known in advance. Thus the term
\begin{equation*}
\rho \bm{D}_{\bm{\eta}_0} \bm{\Sigma}\bm{D}^{T}_{\bm{\eta}_0}    
\end{equation*}
may be viewed as the penalty for substituting estimators for the true values of the parameters. The penalty may also be rewritten as
$\rho\sum_{j=1}^{K} \kappa_{j} \bm{D}_{j} \bm{\Sigma}_{j} \bm{D}^{T}_{j}$ where $\bm{D}_{j} = \mathbb{E}_{0}(\partial\bm{\psi}/\partial{\bm{\eta}_{j})}$ which expresses its dependence on the relative sample sizes, the asymptotic variances of the historical estimates and the sensitivity of the estimation procedure with respect to the historical estimators, embodied in the matrices $\bm{D}_1,\ldots,\bm{D}_K$. 

\begin{remark} \label{rem.rho.small}
Note that if $\rho$ is very small which occurs when $m \gg n$, then the penalty is inconsequential, i.e., that asymptotic variance of $\bar{\bm{\theta}}_{A}$ is close to its variance when $\bm{\eta}_{1},\ldots,\bm{\eta}_{K}$ are fully known. 
\end{remark}

\subsection{Type {\rm II} Problems}

Next, consider the case where both $\bm{\theta}$ and $\bm{\eta}$ are estimable using the data $\mathcal{D}$ observed in the current study $\mathcal{S}$. In this case $\bm{\omega}$ is estimated using a two step procedure. In the first step the data $\mathcal{D}$ is used to obtain the pair $(\tilde{\bm{\theta}},\tilde{\bm{\eta}})^T$ which simultaneously solve 
\begin{equation} \label{Eq.Type.II.Step.1}
\bm{\Psi}(\bm{\theta },\bm{\eta})=\bm{0} \quad \mbox{and} \quad \bm{\Gamma}(\bm{\theta },\bm{\eta})=\bm{0}.  
\end{equation}
The estimating equation $\bm{\Psi}$ is the estimating function for $\bm{\theta}$ for fixed known value of $\bm{\eta}$, as in Type {\rm I} Problems, whereas the estimating function $\bm{\Gamma}$ is the estimating function for $\bm{\eta}$ for fixed value of $\bm{\theta}$; it will not play any role in our developments with the exception of Remark \ref{Rem.More.B&C} appearing in the Appendix. Under standard regularity conditions, cf., the conditions listed just before the statement of Theorem \ref{Thm.Limit.4.Estimator.I}, the estimators $(\tilde{\bm{\theta}},\tilde{\bm{\eta}})^T$ satisfy
\begin{equation} \label{Eq.LST.New}
\sqrt{n}(\tilde{\bm{\theta}}-\bm{\theta}_{0},\tilde{\bm{\eta}}-\bm{\eta}_{0})^T 
\Rightarrow \mathcal{N}_{p+q}(\bm{0},\bm{\Upsilon})
\end{equation}
where $\bm{\Upsilon}$ is assumed to be a non--singular variance matrix which can be consistently estimated from the data by, say $\tilde{\bm{\Upsilon}}$, the standard sandwich estimator (Van der Vaart, 2000). For convenience we may partition $\bm{\Upsilon}$ as
\begin{equation}
\bm{\Upsilon} = 
\begin{pmatrix}
\bm{\Upsilon}_{\bm{\theta}\bm{\theta}}  & \bm{\Upsilon}_{\bm{\theta}\bm{\eta}}\\
\bm{\Upsilon}_{\bm{\eta}\bm{\theta}}  & \bm{\Upsilon}_{\bm{\eta}\bm{\eta}}
\end{pmatrix}
\end{equation}
where $\bm{\Upsilon}_{\bm{\theta}\bm{\theta}}$ and $\bm{\Upsilon}_{\bm{\eta}\bm{\eta}}$ denote the marginal asymptotic variances of $\tilde{\bm{\theta}}$ and $\tilde{\bm{\eta}}$, respectively, and $\bm{\Upsilon}_{\bm{\theta}\bm{\eta}}$ is their asymptotic covariance. Naturally, a similar partition holds for $\tilde{\bm{\Upsilon}}$. Furthermore, as in Section \ref{Sec.3.1}, at our disposal are $K$ independent historical estimators of $\bm{\eta }_{1},\ldots,\bm{\eta}_{K}$ obtained using studies of sizes $m_1,\ldots,m_K$ which satisfy $\sqrt{m_{j}}(\widehat{\bm{\eta }}_{j}-\bm{\eta }_{j,0})\Rightarrow \mathcal{N}_{q_{j}}(\bm{0},\bm{\Sigma }_{j})$, where, again, it is assumed that $\bm{\Sigma }_{j}$ are non--singular and can be consistently estimated for all  $j=1,\ldots,K$. Thus 
\begin{equation} \label{Eq.LST.Historical}
\sqrt{m}(\widehat{\bm{\eta }}-\bm{\eta }_{0})\Rightarrow 
\mathcal{N}_{q}(\bm{0},\bm{\Sigma})
\end{equation}
where $\bm{\Sigma}$ is given in the statement of Theorem \ref{Thm.Limit.4.Estimator.I}. Let $\widehat{\bm{\Sigma}}$ be a consistent estimator of $\bm{\Sigma}$. 

The historic and current estimators of $\bm{\eta}$ can be aggregated, or combined, in many ways. Lemma \ref{Lemma.Estimators.II}, appearing in the Appendix, suggests using the estimator
\begin{equation} \label{Eq.eta.bar}
\bar{\bm{\eta}} = (n\tilde{\bm{\Upsilon}}_{\bm{\eta\eta}}^{-1}+m\widehat{\bm{\Sigma}}^{-1})^{-1} (n\tilde{\bm{\Upsilon}}_{\bm{\eta\eta}}^{-1}\tilde{\bm{\eta}}+m\widehat{\bm{\Sigma}}^{-1}\widehat{\bm{\eta}})    
\end{equation}
which is the MLE under normality assuming that the matrices ${\bm{\Upsilon}}_{\bm{\eta\eta}}$ and $\bm{\Sigma}$ are known. Note that 
$$\bar{\bm{\eta}}= \bm{W}_1\tilde{\bm{\eta}}+\bm{W}_2\widehat{\bm{\eta}}+o_{p}(1)$$
where the weights $\bm{W}_1$ and $\bm{W}_2$ are the symmetric matrices
\begin{equation} \label{Eq.W1&W2}
\bm{W}_1 = (\gamma\bm{\Upsilon}_{\bm{\eta\eta}}^{-1}+(1-\gamma)\bm{\Sigma}^{-1})^{-1}\gamma\bm{\Upsilon}_{\bm{\eta\eta}}^{-1} \quad \textrm{and} \quad \bm{W}_2 = (\gamma\bm{\Upsilon}_{\bm{\eta\eta}}^{-1}+(1-\gamma)\bm{\Sigma}^{-1})^{-1}(1-\gamma)\bm{\Sigma}^{-1}  
\end{equation}
which satisfy $\bm{I} =\bm{W}_1+\bm{W}_2$ with $\gamma = \lim(n/(n+m))$. Thus \eqref{Eq.eta.bar} differs from the best linear unbiased estimator by at most an $o_{p}(1)$ term.  

In the second step we find $\bar{\bm{\theta}}_{B}$ by solving 
\begin{equation} \label{Eq.Type.II.Step.2}
\bm{\Psi}(\bm{\theta },\bar{\bm{\eta}})=0.  
\end{equation}
where $\bar{\bm{\eta}}$ is given by \eqref{Eq.eta.bar}. We now have:

\begin{theorem} \label{Thm.Limit.4.Estimator.IIa}
Let $\bar{\bm{\theta}}_{B}$ be a solution to \eqref{Eq.Type.II.Step.2} where $\bar{\bm{\eta}}_{B}=\bar{\bm{\eta}}$ is given in \eqref{Eq.eta.bar}. Assume that the regularity conditions of Theorem \ref{Thm.Limit.4.Estimator.I} hold. Then
\begin{equation}
\sqrt{n}(\bar{\bm{\theta}}_{B}-\bm{\theta}_{0},\bar{\bm{\eta}}_{B}-\bm{\eta}_{0}) \Rightarrow \mathcal{N}_{p+q}(\bm{0},\bm{B})    
\end{equation}
where
\begin{equation*} \label{Eq.B}
\bm{B} =   
\begin{pmatrix}
\bm{B}_{\bm{\theta\theta}}  & \bm{B}_{\bm{\theta\eta}} \\
\bm{B}_{\bm{\eta\theta}}    & \bm{B}_{\bm{\eta\eta}}
\end{pmatrix}
\end{equation*}
with
\begin{align*}
\bm{B}_{\bm{\theta\theta}}&= (\bm{D}^{-1}_{\bm{\theta}_0})[\bm{\Sigma}_{\bm{\psi}} +  \bm{D}_{\bm{\eta}_0}(\bm{\Upsilon}_{\bm{\eta\eta}}^{-1}+(\rho\bm{\Sigma})^{-1})^{-1}\bm{D}^{T}_{\bm{\eta}_0}]  (\bm{D}^{-1}_{\bm{\theta}_0})^{T},\\
\bm{B}_{\bm{\theta\eta}}&=-\bm{D}_{\bm{\theta}_{0}}^{-1}\bm{D}_{\bm{\eta}_{0}}(\bm{\Upsilon}_{\bm{\eta\eta}}^{-1}+(\rho\bm{\Sigma})^{-1})^{-1},\\
\bm{B}_{\bm{\eta\eta}}&=(\bm{\Upsilon}_{\bm{\eta\eta}}^{-1}+(\rho\bm{\Sigma})^{-1})^{-1}.
\end{align*}
\end{theorem}

Although the mechanics are slightly more involved the proof of Theorem \ref{Thm.Limit.4.Estimator.IIa} builds on the  proof of Theorem \ref{Thm.Limit.4.Estimator.I}. Moreover, the structure of the asymptotic variance matrices $\bm{A}$ and $\bm{B}$ are analogous with the  exception that the variance matrix $\rho\bm{\Sigma}$ appearing in $\bm{A}$ is replaced with $(\bm{\Upsilon}_{\bm{\eta\eta}}^{-1}+(\rho\bm{\Sigma})^{-1})^{-1}$ in $\bm{B}$. 

\begin{remark}
Observe that $(\bm{\Upsilon}_{\bm{\eta\eta}}^{-1}+(\rho\bm{\Sigma})^{-1})^{-1} \rightarrow \bm{0}$ as $\rho \rightarrow 0$ so the conclusions of Remark \ref{rem.rho.small} hold here as well.
\end{remark}

It is clear that whenever the model for $\mathcal{D}$ identifies $\bm{\omega}$ both $(\bar{\bm{\theta}}_{B},\bar{\bm{\eta}}_{B})$ and $(\bar{\bm{\theta}}_{A},\bar{\bm{\eta}}_{A})$ can be computed. Next, using the concept of the Loewner order we show the former is superior to the latter. Recall that the matrix $\bm{V}_1$ is said to be smaller in the Loewner order compared with the matrix $\bm{V}_2$ if $\bm{V}_2-\bm{V}_1$ is non--negative definite (Pukelsheim, 2006). This relationship is denoted by $\bm{V}_1\preceq\bm{V}_2$. Suppose now that $\bm{V}_1$ and $\bm{V}_2$ are the variances of two (asymptotically) unbiased estimators. Then $\bm{V}_1\preceq\bm{V}_2$ implies that the estimator associated with $\bm{V}_1$ is more efficient than the estimator associated with $\bm{V}_2$. This means, for example, that the confidence ellipsoid associated with $\bm{V}_1$ lies within the confidence ellipsoid associated with $\bm{V}_2$.

\begin{theorem} \label{Thm.Order.A&B}
Whenever the data $\mathcal{D}$ identifies $\bm{\omega}$ we have
\begin{equation}
\bm{B} \preceq \bm{A}.
\end{equation}
Moreover, for any function $\bm{\Phi}$ we have $\bm{V}_{\bm{B}}^{\bm{\Phi}} \preceq \bm{V}_{\bm{A}}^{\bm{\Phi}}$ where $\bm{V}_{\bm{A}}^{\bm{\Phi}}$ and $\bm{V}_{\bm{B}}^{\bm{\Phi}}$ are the asymptotic variances of $\bm{\Phi}(\bar{\bm{\theta}}_{A},\bar{\bm{\eta}}_{A})$ and $\bm{\Phi}(\bar{\bm{\theta}}_{B},\bar{\bm{\eta}}_{B})$ respectively. 
\end{theorem}

Theorem \ref{Thm.Order.A&B} indicates that, if possible, it is always asymptotically beneficial to estimate both $\bm{\theta}$ and $\bm{\eta}$ using the data $\mathcal{D}$ collected in the study $\mathcal{S}$. Moreover, Theorem \ref{Thm.Order.A&B} holds also when only a sub--vector of $\bm{\eta}$ is identified by the data $\mathcal{D}$.  

\bigskip

Another variant of Type {\rm II} Problems occurs when the data $\mathcal{D}$ is not available, but nevertheless the estimators $(\tilde{\bm{\theta}},\tilde{\bm{\eta}})$ from the current study as well as their estimated variance, i.e., $\tilde{\bm{\Upsilon}}$ is given. The objective is then to combine the the current estimators \eqref{Eq.LST.New} with the historical estimators \eqref{Eq.LST.Historical}. To this end we propose estimating $\bm{\theta}$ by 
\begin{equation} \label{Eq.theta.bar}
\bar{\bm{\theta}}_{C} = \tilde{\bm{\theta}} - \tilde{\bm{\Upsilon}}_{\bm{\theta}\bm{\eta}}\tilde{\bm{\Upsilon}}_{\bm{\eta}\bm{\eta}}^{-1}(\tilde{\bm{\eta}}-  \bar{\bm{\eta}}_{C})
\end{equation}
where $\bar{\bm{\eta}}_{C}=\bar{\bm{\eta}}$ is given by \eqref{Eq.eta.bar}. The estimators \eqref{Eq.eta.bar} as well as \eqref{Eq.theta.bar} are motivated by Lemma \ref{Lemma.Estimators.II} and Remark \ref{remark.estimators.II} appearing in the Appendix.  
\begin{theorem} \label{Thm.Limit.4.Estimator.IIb}
Let $(\bar{\bm{\theta}}_{C},\bar{\bm{\eta}}_{C})^T$ be defined by \eqref{Eq.eta.bar} and \eqref{Eq.theta.bar}. Suppose further that \eqref{Eq.LST.New} and \eqref{Eq.LST.Historical} hold and both $\bm{\Upsilon}$ and $\bm{\Sigma}$ can be consistently estimated. Then as $n \rightarrow \infty$ we have  
\begin{equation*}
\sqrt{n}(\bar{\bm{\theta}}_{C}-\bm{\theta}_{0},\bar{\bm{\eta}}_{C}-\bm{\eta}_{0})^T \Rightarrow \mathcal{N}_{p+q}(\bm{0},\bm{C})
\end{equation*}
where $\bm{C}=\bm{M}\bm{V}\bm{M}^T$ with
\begin{equation} \label{Eq.V&M}
\bm{V} =   
\begin{pmatrix}
\bm{\Upsilon}  & \bm{0} \\
\bm{0}  & \rho\bm{\Sigma}
\end{pmatrix}
\quad \textrm{and} \quad \bm{M}= 
\begin{pmatrix}
\bm{I}  & -\bm{R}\bm{W}_{2} & \bm{R}\bm{W}_{2} \\
\bm{0}  & \bm{W}_{1}        & \bm{W}_{2}
\end{pmatrix}.
\end{equation}
The matrices $\bm{W}_1$ and $\bm{W}_2$ are defined in \eqref{Eq.W1&W2} and $\bm{R}=\bm{\Upsilon}_{\bm{\theta}\bm{\eta}}\bm{\Upsilon}_{\bm{\eta}\bm{\eta}}^{-1}$. Moreover, we have:  
\begin{align*}
\bm{C}_{\bm{\theta\theta}}&=\bm{\Upsilon}_{\bm{\theta\theta}} - \bm{\Upsilon}_{\bm{\theta\eta}} \bm{\Upsilon}_{\bm{\eta\eta}}^{-1} \bm{W}_2 \bm{\Upsilon}_{\bm{\theta\eta}}^{T},\\
\bm{C}_{\bm{\theta\eta}}&=\bm{\Upsilon}_{\bm{\theta\eta}}\bm{W}_1\\
\bm{C}_{\bm{\eta\eta}}&=(\bm{\Upsilon}_{\bm{\eta\eta}}^{-1}+(\rho \bm{\Sigma})^{-1})^{-1}.
\end{align*}
\end{theorem}

Theorem \ref{Thm.Limit.4.Estimator.IIb} describes the large sample behaviour of the estimators \eqref{Eq.eta.bar} and \eqref{Eq.theta.bar}. Further insight is facilitated by considering the simplest possible situation, i.e.,  when $(\theta,\eta)\in\mathbb{R}^2$, in which case $\sqrt{m}(\widehat{\eta}-\eta_{0})\Rightarrow\mathcal{N}(0,\sigma^2)$ for the historical data, whereas for the current study $\sqrt{n}(\tilde{\theta}-\theta_{0},\tilde{\eta}-\eta_{0})^{T} \Rightarrow\mathcal{N}(0,\bm{\Upsilon})$ where
$$
\bm{\Upsilon}=
\begin{pmatrix}
\upsilon_{\theta\theta}^{2}  & \upsilon_{\theta\eta} \\
\upsilon_{\theta\eta}        & \upsilon_{\eta\eta}^{2}
\end{pmatrix}.
$$
It is not hard to see that \eqref{Eq.theta.bar} reduces to
$\bar{\theta}=\tilde{\theta}-(\tilde{\upsilon}_{\theta\eta}/\tilde{\upsilon}_{\eta\eta}^2)(\tilde{\eta}-\bar{\eta})$ where $\bar{\eta} = w_{1}^{*}\tilde{\eta}+w_{2}^{*}\widehat{\eta}$ with
\begin{equation*}
w_{1}^{*} = \frac{n/\tilde{\upsilon}_{\eta\eta}^2}{n/\tilde{\upsilon}_{\eta\eta}^2+m/\widehat{\sigma}^2} \quad \textrm{and} \quad w_{2}^{*} = \frac{m/\widehat{\sigma}^2}{n/\tilde{\upsilon}_{\eta\eta}^2+m/\widehat{\sigma}^2}.
\end{equation*}
Furthermore $\bm{C}_{\theta\theta}$ simplifies to
\begin{equation} \label{Eq.GammaII.2dim}
\upsilon_{\theta\theta}^2-\frac{\upsilon_{\theta\eta}^2}{\upsilon_{\eta\eta}^2}w_2 =  \upsilon_{\theta\theta}^2(1-w_2r^2),   
\end{equation}
where $r=\upsilon_{\theta\eta}/(\upsilon_{\theta\theta}\upsilon_{\eta\eta})$ is the asymptotic correlation between $\tilde{\theta}$ and $\tilde{\eta}$ and 
$$
w_2 = \frac{ (1-\gamma)/\sigma^{2} }{ \gamma/\upsilon_{\eta\eta}^2 + (1-\gamma)/\sigma^{2} }
$$ 
is the limiting value of $w_{2}^{*}$ as $n/(n+m) \rightarrow \gamma$. It follows that the asymptotic relative efficiency of $\bar{\theta}$ to $\tilde{\theta}$ is $1-w_2r^2$, which is at most unity (when  $\upsilon_{\theta\eta}=0$) and no less than $1-r^2$ (when $\gamma$ is close to $0$). Clearly, the historical estimates are useful only if the covariance $\upsilon_{\theta\eta}$ is non--zero and highly useful whenever $w_2$ is close to unity. A similar but more involved analysis applies when the parameters are multidimensional. 

We emphasize that the structure of the estimators $\bar{\bm{\eta}}_{C}$ and $\bar{\bm{\theta}}_{C}$ as well as the form of $\bm{C}$ are related to, but much more general, than results obtained in the literature on both double sampling and monotone missing normal data (Andersen 1957, Morrison 1971, Kanda and Fujikoshi 1998). Double sampling is a widely used technique in survey sampling, where the estimator is also known as the generalized regression estimator (Thompson, 1997), as well as in other applications, cf. Davidov and Haitovsky (2000), Chen and Chen (2000) and the references therein. We also note that equation \eqref{Eq.GammaII.2dim} is a generalization of the formulas obtained for the usual double sampling estimator (e.g., Tamhane 1978) where $w_2 = m/(n+m)$. The following Theorem substantially generalizes on results obtained in the literature on both the double sampling and monotone missing data. 

\begin{theorem} \label{Thm.Order.C&U}
We have
\begin{equation} \label{Eq.C.less.Upsilon}
\bm{C} \preceq \bm{\Upsilon}   
\end{equation}
Moreover, for any function $\bm{\Phi}$ we have $\bm{V}_{\bm{C}}^{\bm{\Phi}} \preceq \bm{V}_{\bm{\Upsilon}}^{\bm{\Phi}}$ where $\bm{V}_{\bm{C}}^{\bm{\Phi}}$ and $\bm{V}_{\bm{\Upsilon}}^{\bm{\Phi}}$ are the asymptotic variances of $\Phi(\bar{\bm{\theta}}_{C},\bar{\bm{\eta}}_{C})$ and $\Phi(\tilde{\bm{\theta}},\tilde{\bm{\eta}})$ respectively. 
\end{theorem}

In words, the estimator $(\bar{\bm{\theta}}_{C},\bar{\bm{\eta}}_{C})$, incorporating the historical estimates and derived by combining $(\tilde{\bm{\theta}},\tilde{\bm{\eta}})$ and $\widehat{\bm{\eta}}$, is more precise than $(\tilde{\bm{\theta}},\tilde{\bm{\eta}})$, the estimator based only on the current study.   

\begin{remark}
It is also important to emphasize that in finite, typically small samples, the estimator $\bm{C}_{\bm{\theta\theta}}$ may be in fact inferior to $\bm{\Upsilon}_{\bm{\theta\theta}}$. This typically occurs when the "regression matrix" $\bm{R}$, see the statement of Theorem \ref{Thm.Limit.4.Estimator.IIb}, is poorly estimated. This feature has been also recognized in the double sampling literature (Tamhane 1978). 
\end{remark} 

A little algebra shows that 
\begin{align*}
\bm{C}_{\bm{\theta\theta}}&=\bm{\Upsilon}_{\bm{\theta\theta}} - \bm{\Upsilon}_{\bm{\theta\eta}} \bm{\Upsilon}_{\bm{\eta\eta}}^{-1} (\bm{\Upsilon}_{\bm{\eta\eta}}^{-1}+(\rho \bm{\Sigma})^{-1})^{-1} (\rho \bm{\Sigma})^{-1} \bm{\Upsilon}_{\bm{\theta\eta}}^{T},\\
\bm{C}_{\bm{\theta\eta}}&=\bm{\Upsilon}_{\bm{\theta\eta}}\bm{\Upsilon}_{\bm{\eta\eta}}^{-1}(\bm{\Upsilon}_{\bm{\eta\eta}}^{-1}+(\rho \bm{\Sigma})^{-1})^{-1}\\
\bm{C}_{\bm{\eta\eta}}&=(\bm{\Upsilon}_{\bm{\eta\eta}}^{-1}+(\rho \bm{\Sigma})^{-1})^{-1}.
\end{align*}
so we can remove the dependence of $\bm{C}$ on the matrices $\bm{W}_1$ and $\bm{W}_2$. 

Clearly, whenever the data $\mathcal{D}$ is available both $(\bar{\bm{\theta}}_{B},\bar{\bm{\eta}}_{B})$ and $(\bar{\bm{\theta}}_{C},\bar{\bm{\eta}}_{C})$ can be calculated where $\bar{\bm{\eta}}_{B}=\bar{\bm{\eta}}_{C}$ are given in \eqref{Eq.eta.bar}. Recall that $\bar{\bm{\theta}}_{B}$ solves $\bm{\Psi}(\bm{\theta},\bar{\bm{\eta}})=\bm{0}$ where $\bm{\Psi}(\bm{\theta},\bm{\eta})=n^{-1}\sum_{i=1}^{n} \bm{\psi}(\bm{\theta},\bm{\eta},\bm{Y}_i)$. Similarly, we can view $\bar{\bm{\theta}}_{C}$ as a solution to (some) estimating equation $\bm{\Lambda}(\bm{\theta},\bar{\bm{\eta}})=\bm{0}$ where $\bm{\Lambda}(\bm{\theta},\bm{\eta})=n^{-1}\sum_{i=1}^{n} \bm{\lambda}(\bm{\theta},\bm{\eta},\bm{Y}_i)$. The form of $\bm{\Lambda}$ can be easily deduced from Lemma \ref{Lemma.Estimators.II} and that of $\bm{\lambda}$ by plugging in the influence functions for $\tilde{\bm{\theta}}$ and $\tilde{\bm{\eta}}$ into $\bm{\Lambda}$. In fact, the precise form of the influence function of $\bar{\bm{\theta}}_{C}$ is readily derived, for more details see Remark \ref{Rem.More.B&C} appearing in the Appendix. It is worth noting that $\bm{\Psi}$ operates on the full data $\mathcal{D}$ whereas $\bm{\Lambda}$ operates on functions thereof namely the estimators $(\tilde{\bm{\theta}},\tilde{\bm{\eta}})$ and $\widehat{\bm{\eta}}$. Thus $(\bar{\bm{\theta}}_{C},\bar{\bm{\eta}}_{C})$ can be viewed as functions of a coarsening of the data $\mathcal{D}$ and therefore is expected to be less efficient than $(\bar{\bm{\theta}}_{B},\bar{\bm{\eta}}_{B})$. This indeed is the case under mild regularity conditions. A formal statement requires the introduction of some additional notation. Let $\bm{h}=\bm{h}(\bm{\theta},\bm{\eta},\bm{Y})$ denote any estimating function and denote $\bm{D}_{\bm{\theta}_{0}}(\bm{h}) = \mathbb{E}_{0}(\partial\bm{h}/\partial{\bm{\theta})}$ and $\bm{D}_{\bm{\eta}_{0}}(\bm{h}) = \mathbb{E}_{0}(\partial\bm{h}/\partial{\bm{\theta})}$. Note that earlier we referred to $\bm{D}_{\bm{\theta}_{0}}(\bm{\psi})$ and $\bm{D}_{\bm{\eta}_{0}}(\bm{\psi})$ simply as $\bm{D}_{\bm{\theta}_{0}}$  and $\bm{D}_{\bm{\eta}_{0}}$. Now:

\begin{theorem} \label{Thm.Order.B&C}
Suppose that the both $\bar{\bm{\omega}}_{B}$ and $\bar{\bm{\omega}}_{C}$ can be obtained. If 
\begin{equation} \label{Eq.Sensitive}
\bm{D}_{\bm{\theta}_{0}}(\bm{\psi})^{-1}\bm{D}_{\bm{\eta}_{0}}(\bm{\psi}) \le  \bm{D}_{\bm{\theta}_{0}}(\bm{\lambda})^{-1}  \bm{D}_{\bm{\eta}_{0}}(\bm{\lambda})
\end{equation} 
component--wise and
\begin{equation} \label{Eq.Efficient} 
(\bm{D}_{\bm{\theta}_{0}}(\bm{\psi})^{-1})\mathbb{E}_{0}(\bm{\psi\psi}^T)(\bm{D}_{\bm{\theta}_{0}}(\bm{\psi})^{-1})^T \preceq (\bm{D}_{\bm{\theta}_{0}}(\bm{\lambda})^{-1})\mathbb{E}_{0}(\bm{\lambda\lambda}^T)(\bm{D}_{\bm{\theta}_{0}}(\bm{\lambda})^{-1})^T   
\end{equation}
in the Loewner order, then  
\begin{equation}
\bm{B} \preceq \bm{C}.
\end{equation}
Moreover, for any function $\bm{\Phi}$ we have $\bm{V}_{\bm{B}}^{\bm{\Phi}} \preceq \bm{V}_{\bm{C}}^{\bm{\Phi}}$ where $\bm{V}_{\bm{B}}^{\bm{\Phi}}$ and $\bm{V}_{\bm{C}}^{\bm{\Phi}}$ are the asymptotic variances of $\bm{\Phi}(\bar{\bm{\theta}}_{B},\bar{\bm{\eta}}_{B})$ and $\bm{\Phi}(\bar{\bm{\theta}}_{C},\bar{\bm{\eta}}_{C})$ respectively. 
\end{theorem}

Condition \eqref{Eq.Efficient} holds when the estimating equation $\bm{\Psi}(\bm{\theta},\bm{\eta}_0)=\bm{0}$ results in more efficient estimators for $\bm{\theta}$ than those resulting from $\bm{\Lambda}(\bm{\theta},\bm{\eta}_0)=\bm{0}$ when $\bm{\eta}=\bm{\eta}_{0}$ is set to its true value. This condition holds for any sensibly choice of $\bm{\Psi}$. In particular it holds for the score equations associated with maximum likelihood estimation. Condition \eqref{Eq.Sensitive} roughly means that $\bm{\Psi}$ is less sensitive to small perturbations in both $\bm{\theta}$ and $\bm{\eta}$ compared with $\bm{\Lambda}$. Conditions \eqref{Eq.Sensitive} are \eqref{Eq.Efficient} are not necessary. For example, the conclusion of Theorem \ref{Thm.Order.B&C} may hold if $\bm{\psi}$ is more sensitive to small perturbations but at the same time much more efficient. We believe that the aforementioned conditions hold broadly and the estimators  $(\bar{\bm{\theta}}_{C},\bar{\bm{\eta}}_{C})$, described in Theorem \ref{Thm.Limit.4.Estimator.IIb}, is generally less efficient than $(\bar{\bm{\theta}}_{B},\bar{\bm{\eta}}_{B})$, described in Theorem \ref{Thm.Limit.4.Estimator.IIa}. For an additional discussion see Remark \ref{Rem.More.B&C} in the Appendix. 

There are, however, situations in which $\bm{B}=\bm{C}$ and situations where $\bar{\bm{\omega}}_{B} =\bar{\bm{\omega}}_{C}$ for any data $\mathcal{D}$. As we shall see in the next section this is the case in normal linear models in which the estimators $(\tilde{\bm{\theta}},\tilde{\bm{\eta}})$ and $\widehat{\bm{\eta}}$ are actually sufficient statistics. Finally, it is worth noting that if $\bm{\Upsilon}_{\bm{\theta\eta}} = \bm{0}$ then the estimator $(\bar{\bm{\theta}}_{C},\bar{\bm{\eta}}_{C})$ does not improve $\tilde{\bm{\theta}}$ whereas there is always an improvement when the full data $\mathcal{D}$ is available. 

\section{Illustrations, applications and numerical results} \label{Sec.Application}

In this section two applications are discussed in detail. In Section \ref{Sec.Application.ANOVA} the two--way ANOVA problem introduced in Section \ref{Sec.Notation.Formulation} is investigated. In particular, various design options for the current study $\mathcal{S}$ are evaluated. It is worth noting that although the abovementioned ANOVA problem is among the simplest possible, its analysis is far from trivial. Next, in Section \ref{Sec.Application.Bliss} we discuss the use of historical estimates in the design of drug interaction studies in the context of Bliss independence. A simple algorithm for the design of such studies is proposed. 

\subsection{Two way ANOVA} \label{Sec.Application.ANOVA}

Recall the ANOVA model of Section \ref{Sec.Notation.Formulation} where the studies $\mathcal{S}_1$ and $\mathcal{S}_2$ were designed to estimate $\bm{\eta}_{1}=(\eta_0,\eta_1)^T$ and $\bm{\eta}_{2}=(\eta_0,\eta_2)^T$, respectively. Note that the parameter $\eta_0$ is estimated in both studies so $\bm{\eta}_{1}$ and $\bm{\eta}_{2}$ are not distinct. Therefore employing any of the aforementioned findings requires the aggregation of the historical estimates as if they came from a single experiment. The historical studies result in the estimates $(\widehat{\eta}_{0}(\mathcal{S}_{1}),\widehat{\eta}_{1}(\mathcal{S}_{1}))$ and  $(\widehat{\eta}_{0}(\mathcal{S}_{2}),\widehat{\eta}_{2}(\mathcal{S}_{2}))$ as well as their standard errors, we can easily back calculate the potentially unobserved sufficient statistics and sample sizes in the studies $\mathcal{S}_1$ and $\mathcal{S}_2$ and estimate $(\eta_0,\eta_1,\eta_2)$ by:
\begin{equation} \label{Eq.eta.ANOVA}
\widehat{\eta}_{1} = \bar{Y}_1(\mathcal{S}_1) - \widehat{\eta}_{0}, \quad \widehat{\eta}_{2} = \bar{Y}_2(\mathcal{S}_2) - \widehat{\eta}_{0}, \quad \mbox{and} \quad 
\widehat{\eta}_{0} = \frac{ m_{1,0}\bar{Y}_0(\mathcal{S}_1) + m_{2,0}\bar{Y}_0(\mathcal{S}_2) }{m_{1,0}+m_{2,0}},
\end{equation}
where the quantity $\bar{Y}_j(\mathcal{S}_i)$ is the average response on treatment $j\in\{0,1,2\}$ in study $i\in\{1,2\}$. Similarly $m_{i,j}$ is the size of of treatment group $j$ in study $i$. It follows, under the usual conditions, that
\begin{equation*}
\sqrt{m}(\widehat{\eta_0}-{\eta }_{0},\widehat{\eta_1}-{\eta }_{1}, 
\widehat{\eta_2}-{\eta }_{2})^T \Rightarrow \mathcal{N}(\boldsymbol{0},\boldsymbol{\Sigma}),
\end{equation*}
for some matrix $\bm{\Sigma}$. Furthermore, if \eqref{eq.twoway.anova} is homoscedastic model with variance $\sigma^2$ and $m_{1,0}=m_{1,1}=m_{2,0}=m_{2,2}$, i.e., the studies $\mathcal{S}_{1}$ and $\mathcal{S}_{2}$ are balanced and of the same size, then it is easy to see that
$$
\bm{\Sigma} =\sigma^2 
\begin{pmatrix}
2  & -2 & -2 \\
-2 & 6  &  2 \\
-2 & 2  &  6
\end{pmatrix}.
$$

\medskip

We will now investigate various designs for a new study $\mathcal{S}$. If the primary focus of $\mathcal{S}$ is inference on $\theta$ then it may, in some circumstances, be advantageous to allocate all $n$ observations to the treatment arm receiving both treatments one and two, i.e., $T_{1}=T_{2}=1$ for all observations. This is clearly a Type {\rm I} problem since $\bm{\omega}$ is not identifiable from $\mathcal{D}$ but given $\bm{\eta}$ the parameter $\theta$ is estimable. Note that an unbiased estimate for $\theta$ is
\begin{equation} \label{Eq.theta.bar.A.anova}
\bar{\theta}_{A} =\bar{Y}_{12}(\mathcal{S})-(\widehat{\eta}_{0}+\widehat{\eta}_{1}+\widehat{\eta}_{2})    
\end{equation}
and it is not hard to see that \eqref{Eq.theta.bar.A.anova} solves \eqref{Eq.Type.I} when  $\psi(\theta,\eta_0,\eta_1,\eta_2,Y_i)=Y_i-\eta_0-\eta_1-\eta_2-\theta$. Thus, $\bm{\Sigma}_{\psi} = \sigma^2$, $\bm{D}_{\theta_{0}} = 1$ and $\bm{D}_{\eta_{0}}= -(1,1,1)$ and it follows that $\bm{A}_{\theta\theta}$, the asymptotic variance of \eqref{Eq.theta.bar.A.anova} as described in Theorem \ref{Thm.Limit.4.Estimator.I}, reduces to 
$$
\sigma^2 \times (1+10\rho) \quad \mbox{where} \quad \rho =\lim \frac{n}{m}.
$$

The second term appearing in the parentheses in the above display is an inflation factor, i.e., the price to pay for substituting estimators for the unknown value of $(\eta_0,\eta_1,\eta_2)$. Note that when $n/m \rightarrow 0$ as both $m\rightarrow \infty$ and $n\rightarrow \infty$ the asymptotic variance of $\bar{\theta}_{A}$  approaches $\sigma^2$. In practice this requires a large current study and even larger historical data. Incidentally, since $\bar{\theta}_{A}$ is a linear function of $\bar{Y}_{12}(\mathcal{S})$ and $(\widehat{\eta}_{0},\widehat{\eta}_{1},\widehat{\eta}_{2})$ it is not hard to see that its exact variance is $\sigma^2(1/n+10/m)$ which coincides with the asymptotic form. 

\medskip

Alternatively, suppose that the study $\mathcal{S}$ allocates $n/4$ observations to all treatment combinations. In this case the data $\mathcal{D}$ identifies $\bm{\omega} = (\theta,\eta_0,\eta_1,\eta_2)^T$, so this is a Type {\rm II} problem. The usual estimators for this design are $\tilde{\eta_{0}} =\bar{Y}_{0}(\mathcal{S})$, $\tilde{\eta_{1}} =\bar{Y}_{1}(\mathcal{S})-\bar{Y}_{0}(\mathcal{S})$, $\tilde{\eta_{2}} =\bar{Y}_{2}(\mathcal{S})-\bar{Y}_{0}(\mathcal{S})$ and
$$
\tilde{\theta} =\bar{Y}_{12}(\mathcal{S})-(\bar{Y}_{1}(\mathcal{S})+\bar{Y}_{2}(\mathcal{S}))+\bar{Y}_{0}(\mathcal{S})
$$
and thus the limiting variance of $(\tilde{\theta},\tilde{\bm{\eta}})^{T}$ is 
$$
\bm{\Upsilon} =\sigma^2 
\begin{pmatrix}
16  & 4 & -8 & -8 \\
4 & 4   & -4 & -4 \\
-8 & -4   & 8 & 4 \\
-8 & -4   & 4 & 8
\end{pmatrix}.
$$
Next we aggregate the historical and current estimators for $\bm{\eta}$. As in Section \ref{Sec.Results} we estimate $\bm{\eta}$ by $\bar{\bm{\eta}} = \bm{W_1}\tilde{\bm{\eta}}+\bm{W_2} \widehat{\bm{\eta}}$ where 
\begin{equation*}
\bm{W_1}^{*} = (n{\bm{\Upsilon}}_{\bm{\eta\eta}}^{-1}+m{\bm{\Sigma}}^{-1})^{-1} n{\bm{\Upsilon}}_{\bm{\eta\eta}}^{-1} \quad \mbox{and} \quad \bm{W_2}{*}= (n\bm{\Upsilon}_{\bm{\eta\eta}}^{-1}+m\bm{\Sigma}^{-1})^{-1} m\bm{\Sigma}^{-1}.
\end{equation*}
Note that the weight matrices are functions of the variances $\bm{\Upsilon}_{\bm{\eta\eta}}$ and $\bm{\Sigma}$ as well as the ratio $n/(n+m)$. Since $\mathcal{D}$ is fully available to us then we can estimate $\theta$ by  
\begin{equation} \label{Eq.theta.bar.B.anova}
\bar{\theta}_{B} =\bar{Y}_{12}(\mathcal{S})-(\bar{\eta}_{0}+\bar{\eta}_{1}+\bar{\eta}_{2}).    
\end{equation}
Note that the estimators \eqref{Eq.theta.bar.A.anova} and \eqref{Eq.theta.bar.B.anova}
are of the same functional form. Further note that the statistic $\bar{Y}_{12}$ in \eqref{Eq.theta.bar.B.anova} is a function of the $n_{12}$ observations $Y_1,\ldots,Y_{n_{12}}$ receiving the treatment combination $T_1=T_2=1$. A straightforward calculation shows that $\bm{B}_{\theta\theta}$ is given by
$$
\sigma^2 \times (\xi_{11}^{-1}+\bm{1}^{T}(\bm{\Upsilon}_{\bm{\eta\eta}}^{-1}+(\rho\bm{\Sigma})^{-1})^{-1}\bm{1}) \quad \mbox{where} \quad \rho = \lim\frac{n}{m},
$$
where $\xi_{11}$ is the fraction of the observations which are assigned to receive both treatments. In situations where the full data is not available to us but $(\tilde{\theta},\tilde{\eta}_0,\tilde{\eta}_1,\tilde{\eta}_2)$ are known we may estimate $\theta$ by $ \bar{\theta}_{C} =\tilde{\theta}-\bm{\Upsilon}_{\theta\bm{\eta}}\bm{\Upsilon}_{\bm{\eta}\bm{\eta}}^{-1}(\tilde{\bm{\eta}}-\bar{\bm{\eta}})$. It can be verified that in this application, in which a normal linear model is involved and all estimators are functions of sufficient statistics, the estimators   $\bar{\theta}_{B}$ and $\bar{\theta}_{C}$ coincide. Therefore $\bar{\theta}_{C}$ is not discussed any further. 

Table \ref{Tab.A.B} provides a comparison of the asymptotic variances of \eqref{Eq.theta.bar.A.anova} and \eqref{Eq.theta.bar.B.anova} for a range of values of $m$ and $n$. 

\medskip
\begin{center}
Table \ref{Tab.A.B} Comes Here.   
\end{center}
\medskip

Table \ref{Tab.A.B} displays asymptotic variances; the variances themselves are found by dividing any entry in the table by the size of the current study in the relevant row. Observe that both $\bm{A}_{\theta\theta}$ and $\bm{B}_{\theta\theta}$ decrease as a function of $m$ for any fixed value of $n$ and increase in $n$ for any fixed $m$. For example when $n=m=100$ 
$\bm{A}_{\theta\theta}=11$ and $\bm{B}_{\theta\theta}=9.3$ whereas when  $m=100$ and $n=5000$ then $\bm{A}_{\theta\theta}=501$ and $\bm{B}_{\theta\theta}=15.69$ and when $m=5000$ and $n=100$ then $\bm{A}_{\theta\theta}=1.2$ and $\bm{B}_{\theta\theta}=4.2$. Thus going down the first column of Table \ref{Tab.A.B} the asymptotic variance $\bm{A}_{\theta\theta}$ is increases by a factor of approximately $45$ whereas that of $\bm{B}_{\theta\theta}$ by the much more modest $1.4$. Similarly going across the first row the asymptotic variances of  $\bm{A}_{\theta\theta}$ and $\bm{B}_{\theta\theta}$ are reduced by a factor of $9.2$ and $2.2$ respectively. Each pair $(n,m)$ provides a direct comparison between the two designs (design $A$, say, which assigns all experimental units in the current study to receive both treatments and design $B$, say, which is a balanced design). Clearly, design $A$ seems preferable in situations where $m$ is much larger that $n$, otherwise design $B$ is to be preferred.  

\bigskip

We now look a bit deeper into the question of optimal design. Suppose, that as before the historical sample of size $m$ where $m_{1,0}=m_{1,1}=m_{2,0}=m_{2,2}$. The objective is to design a study $\mathcal{S}$ of size $n$ which would minimize the variance of the estimate of $\theta$. Let $\xi_{ij}$ denote the proportion of observation who received treatment combination $i \times j$ where $i$ and $j$ are in $\{0,1\}$. Thus the design vector is nothing but $\bm{\xi}=(\xi_{00},\xi_{10},\xi_{01},\xi_{11})$. For simplicity we will assume an approximate design which implies that $\bm{\xi}$ lies in the unit simplex. Moreover, estimating $\theta$ requires that $\xi_{11}>0$. Note that Table \ref{Tab.A.B} considers only designs with $\bm{\xi}=(0,0,0,1)$ and $\bm{\xi}=(1/4,1/4,1/4,1/4)$. It is not hard to see that in this relatively simple setting the optimal design (in the interior of the simplex) is attained when $\xi_{11}^{-1}+\bm{1}^{T}(\bm{\Upsilon}_{\bm{\eta\eta}}^{-1}+(\rho\bm{\Sigma})^{-1})^{-1}\bm{1}$ is minimized where 
\begin{equation*}
\bm{\Upsilon} =\sigma^2 
\begin{pmatrix}
\frac{1}{\xi_{00}}+\frac{1}{\xi_{10}}+\frac{1}{\xi_{01}}+\frac{1}{\xi_{11}}  & \frac{1}{\xi_{00}} & -(\frac{1}{\xi_{00}}+\frac{1}{\xi_{10}}) & -(\frac{1}{\xi_{00}}+\frac{1}{\xi_{01}}) \\
\frac{1}{\xi_{00}} & \frac{1}{\xi_{00}}   & -\frac{1}{\xi_{00}} & -\frac{1}{\xi_{00}} \\
-(\frac{1}{\xi_{00}}+\frac{1}{\xi_{10}}) & -\frac{1}{\xi_{00}} & (\frac{1}{\xi_{00}}+\frac{1}{\xi_{10}}) & \frac{1}{\xi_{00}} \\
-(\frac{1}{\xi_{00}}+\frac{1}{\xi_{01}}) & -\frac{1}{\xi_{00}}   & \frac{1}{\xi_{00}} & \frac{1}{\xi_{00}}+\frac{1}{\xi_{10}}
\end{pmatrix}.    
\end{equation*}
Symmetry consideration imply that under optimality $\xi_{01}=\xi_{10}$ and since $\xi_{00}=1-2\xi_{10}-\xi_{11}$ the minimization involves only a two dimensional search. Table \ref{Tab.Opt} provides the optimal design, i.e., the vector $\bm{\xi}$ for estimating $\theta$ for various values of the ratio $\rho = n/m$ found by a grid search with step $0.001$ and the restriction that $\xi_{00} \ge 0.02$. This restriction is necessary; otherwise the matrix $\bm{\Upsilon}$ can not be inverted.

\medskip
\begin{center}
Table \ref{Tab.Opt} Comes Here.   
\end{center}
\medskip

The designs appearing in Table \ref{Tab.Opt} are generally superior to those in Table \ref{Tab.A.B}. For example when $\rho=1$ we find that the asymptotic variances in Table \ref{Tab.A.B} are $11.0$ and $9.3$ whereas the corresponding optimal asymptotic variance given in Table \ref{Tab.Opt} is $8.00$. Further note that for large $\rho$, i.e., when $n$ is larger than $m$, we find that $\xi_{01}=\xi_{10}=1/4$ and that the difference between $\xi_{11}$ and $\xi_{00}$ decreases in $\rho$. We believe that the balanced design is optimal when $\rho \rightarrow \infty$. 
Also note that when $\rho$ is smaller than a $1/4$, i.e., when $n$ is relatively small to $m$, then $\xi_{00}=0.02$, and $\xi_{01}=\xi_{10}=0.01$ which are the smallest possible values allowed by our algorithm. This suggest that further minor improvements are possible by setting $\xi_{00}=0$ and/or $\xi_{01}=\xi_{10}=0$. Clearly when $\xi_{01}=\xi_{10}=\xi_{00}$ we have a Type {\rm I} Problem. 

Therefore we next consider the situation that $\xi_{00}=0$ and $\xi_{01}=\xi_{10}>0$, in which case the current study comprises of three groups and thus three group means: $\bar{Y}_{1}(\mathcal{S})$, $\bar{Y}_{2}(\mathcal{S})$ and $\bar{Y}_{12}(\mathcal{S})$. It is important to note that with these data alone we can not estimate $\bm{\omega}$. Nevertheless, the pair $(\bar{Y}_{1}(\mathcal{S}),\bar{Y}_{2}(\mathcal{S}))^{T}$ whose mean is $(\eta_{0}+\eta_{1},\eta_{0}+\eta_{2})$ can be aggregated with with $\widehat{\bm{\eta}}$ the historical estimate of $\bm{\eta}$. By an appropriate modification of Lemma \ref{Lemma.Estimators.II} it can be shown that $\bm{\eta}$ can be estimated by
\begin{equation} \label{Eq.eta.dagger}
\bm{\eta}^{\dagger}=(n\bm{A}^T\bm{V}^{-1}\bm{A}+m\bm{\Sigma}^{-1})^{-1}(n\bm{A}^T\bm{V}^{-1}\bm{S}+m\bm{\Sigma}^{-1}\widehat{\bm{\eta}})    
\end{equation}
where $\bm{S}=(\bar{Y}_{1}(\mathcal{S}),\bar{Y}_{2}(\mathcal{S}))^{T}$, $\bm{V} = \sigma^{2}{\rm diag}(\xi_{01}^{-1},\xi_{10}^{-1})$ is its asymptotic variance and 
$$
\bm{A} = 
\begin{pmatrix}
1  & 1 & 0 \\
1 &  0 & 1 
\end{pmatrix}.
$$
is the matrix which satisfies $\mathbb{E}(\bm{S})=\bm{A\eta}$. Note that \eqref{Eq.eta.dagger} is of the same form as \eqref{Eq.eta.bar} but with $\bm{A^{T}V^{-1}A}$ instead of $\bm{\Upsilon}_{\bm{\eta\eta}}$. Now, let $\bar{\theta}_{D}$ denote the solution to $\bm{\Psi}(\theta,\bm{\eta}^{\dagger})=0$ which is nothing but 
\begin{equation} \label{Eq.theta.bar.D.anova}
\bar{\theta}_{D} =\bar{Y}_{12}(\mathcal{S})-(\eta_{0}^{\dagger}+\eta_{1}^{\dagger}+\eta_{2}^{\dagger})    
\end{equation}
A straightforward calculation shows that the asymptotic variance of $\bar{\theta}_{D}$ is given by
$$
\sigma^2 \times (\xi_{11}^{-1}+\bm{1}^{T}(\bm{A^{T}V^{-1}A}+(\rho\bm{\Sigma})^{-1})^{-1}\bm{1}) \quad \mbox{where} \quad \rho =\lim \frac{n}{m}.
$$
The formula above is useful in finding the optimal design for small values of $\rho$ when $\xi_{00}=0$. For example when $\rho=1/8$ then the design $\bm{\xi} = (0,0.0005,0.0005, 0.9990)$ results in a variance of $2.25$ (actually $2.250751$) which is slightly smaller than $2.27$ the reported variance in the first row of Table \ref{Tab.Opt}. Finally we note that when $\rho=1/8$ then $A_{\theta\theta}$ equals (precisely) $2.25$ which means that in this application a design for Type {\rm I} would be the most effective.

\subsection{Using historical estimates in drug interaction studies} \label{Sec.Application.Bliss}

This subsection deals with the optimal design of drug interaction studies. Consider two drugs $\bm{D}_{1}$ and $\bm{D}_2$ with no--effect probabilities $\eta_1$ and $\eta_2$, respectively and let $\theta$ denote the no--effect probability when both drugs are administered together. The drugs are called Bliss independent, see, Bliss (1939), Liu et al. (2018), if
\begin{equation}\label{bliss}
\theta=\eta_{1}\eta_{2}.
\end{equation}
If \eqref{bliss} does not hold and $\theta<\eta_{1}\eta_{2}$ there is synergy among the drugs, otherwise there is antagonism. The concept of Bliss independence has seen a recent resurgence of interest as the need to asses the benefit of combination therapies and drug--drug interactions has increased. Some current references are Pallmann and Schaarschmidt (2016), Palmer and Sorger (2017), Russ and Kishony (2018), Qin et al. (2018) and Niu et al (2019). Drug interaction studies are often carried out as single--dose experiment, e.g., Ansari et al. (2008), where the interaction is assessed by considering a single dose of each of the two drugs. A more elaborate design, which we will not consider here, assesses multiple drugs and doses using response surface methodology as in Lee (2010). 

Naturally, the quantity of interest in drug interaction studies is
\begin{equation}\label{logbliss}
\Phi(\theta,\bm{\eta})=\log(\theta)-\log(\eta_{1})-\log(\eta_{2}).
\end{equation}
The formulation in (\ref{logbliss}) links the problem discussed here to the ANOVA setup considered earlier. In many applications of single dose interaction tests, whether using historical data or not,  an explicit or implicit asymptotic argument is used, and the theoretical results for the asymptotic case presented above are relevant. For example,  Demidenko and Miller (2019) describes a Daphnia acute test with two stressors, single doses of CuSO4 and of NiCl, where the numbers of surviving organisms in water were counted after 48 hours. The observations reported were the surviving fractions of organisms only, without reporting their original numbers thus, essentially, assuming their original numbers were very high, i.e., applying an asymptotic argument. But as pointed out by Pallmann and Schaarschmidt (2016), in  single-dose experiments, correct statistical analysis should rely on the observed frequencies, and not on the observed rates of success or failure. Therefore the sample sizes used in each arm of the experiment are of crucial importance and in this subsection we provide finite sample results.

For simplicity suppose that there exists historical estimates of $\eta_{1}$ and $\eta_{1}$ based on independent binomial experiments with sizes $m_1$ and $m_2$. Suppose further that current study allows for the recruitment of $n$ experimental units, $n_1$ of which will receive $\bm{D}_1$, $n_2$ will receive $\bm{D}_2$ and $n_{12}$ will receive both drugs. Obviously 
\begin{equation} \label{Eq.Bliss.Constriant}
n=n_{1}+n_{2}+n_{12}    
\end{equation}
and $\theta$ can not be estimated unless $n_{12}>0$. However it is possible that $n_{1}=n_{2}=0$. The goal is to allocate the experimental units optimally, which is equivalent to the problem of optimally allocating $n+m_1+m_2$ observations in an experiment in which the single dose arms are no smaller than $m_1$ and $m_2$, respectively. The optimal design problem can be approximated as the minimization of the large sample variance of of \eqref{logbliss} 
\begin{equation} \label{bliss.var}
\frac{1}{n_{12}}\frac{1-\theta}{\theta}+\frac{1}{n_1+m_1}\frac{1-\eta_{1}}{\eta_1}+\frac{1}{n_2+m_2}\frac{1-\eta_2}{\eta_2},    
\end{equation}
subject to the constraint \eqref{Eq.Bliss.Constriant}. 

In contrast with the design problem encountered in Section \ref{Sec.Application.ANOVA} the design criterion depends on the unknown parameters, i.e., the probabilities $\theta$ and $\bm{\eta}$. We propose allocating observations as if $\eta_1=\widehat{\eta_1}$, $\eta_2=\widehat{\eta_2}$ and $\theta = \widehat{\eta_1}\widehat{\eta_2}$ is equal to its estimated value under the hypothesis of Bliss independence. 
 
One simple approach to the minimization of \eqref{bliss.var} is the following greedy iterative procedure, which sequentially allocates observations into the condition where the variance is reduced most.

\medskip

\vspace{2mm}



\vspace{2mm}

ALGORITHM

\begin{algorithmic}
\STATE
 $n_{12} \gets 1,\,\, n_{1} \gets 0, \,\, n_{2} \gets 0$
\IF {$n=n_{1}+n_{2}+n_{12}$}
\STATE stop
\ELSE 
\STATE $R_{12}(n_{12}) \gets \frac{1}{n_{12}}
\frac{1-\widehat{\eta}_{1}\widehat{\eta}_{2}}{\widehat{\eta}_{1}\widehat{\eta}_{2}}$ 
\STATE $R_1(n_1) \gets \frac{1}{n_{1}+m_{1}}\frac{1-\widehat{\eta}_{1}}{\widehat{\eta}_{1}}$
\STATE $R_2(n_2) \gets\frac{1}{n_{2}+m_{2}}\frac{1-\widehat{\eta}_{2}}{\widehat{\eta}_{2}}$
\STATE $C \gets min\{R_{12}(n_{12}+1)-R_{12}(n_{12}), R_{1}(n_{1}+1)-R_{1}(n_{1}), R_{2}(n_{2}+1)-R_{2}(n_{2})\}$
\IF {$C=R_{12}(n_{12}+1)-R_{12}(n_{12})$}
\STATE $n_{12} \gets n_{12}+1$
\ENDIF
\IF {$C=R_{1}(n_{1}+1)-R_{1}(n_{1})$}
\STATE $n_{1} \gets n_{1}+1$
\ENDIF
\IF {$C=R_{2}(n_{2}+1)-R_{2}(n_{2})$}
\STATE $n_{2} \gets n_{2}+1$
\ENDIF
\ENDIF
\end{algorithmic}
\vspace{2mm}
\vspace{5mm}

For example, if $m_1=30$, $m_1=50$, $\widehat{\eta}_{1}=0.7$ and $\widehat{\eta}_{2}=0.8$, the first $55$ observations would be put in the arm where both treatments administered, before the 56th observation would be used to improve the estimate of $\hat{\eta}_{1}$. Table \ref{t1} contains a tabulation of the optimal allocation of $(n_{12}, n_{1}, n_{2})$. For selected combinations of the  values of  $m_1$, $m_2$, $\eta_1$, $\eta_2$ the table gives the minimal value of $n$, denoted as $n_{{\rm min}}$, for which  replications of the historic observations is needed, and then the optimal allocation for $n_{{\rm min}}$. As one would expect, when $\theta$ is closer to $0.5$ than $\eta_1$ or $\eta_2$, a larger sample size $n_{12}$ is allocated in the optimal design to estimating $\theta$, than $m_1$ or $m_2$. In the opposite case, $n_{12}$ is smaller than $m_1$ or $m_2$.  

\medskip
\begin{center}
Table \ref{t1} Comes Here.   
\end{center}
\medskip

\section{Summary and discussion} \label{Sec.Disscusion}

Historical findings often inspire current research whether or not they formally incorporate historical data or estimates. Even when historical data or estimates are explicitly incorporated in the analysis, which is quite common in practice, the variability of these estimates is rarely properly accounted for in the analysis. Relying on historical estimates is particularly important when they are essential for model fitting but impossible, or very expensive, to obtain in the context of a current study. A partial list of examples, drawn from the scientific literature, was furnished earlier; many more exist. However, it is very difficult to find published research where the details are given to the extent which would make the replication of the analysis possible. This limits one's ability to apply the results of this paper to published research. However, the results presented here will inform future researchers of the scope and use of historical estimates and provide a tool kit for doing so. We also hope that our investigation may have an effect on publication standards.

Different disciplines exhibit different modes of using historical data. Social scientists often incorporate estimates from surveys in the process of model fitting, whereas biologists and engineers may use parameters estimated in experiments which are very different than their own. One way, of course, of incorporating historical estimates is using prior distributions within the Bayesian framework. For recent examples see Hoff (2019) and Bryan and Hoff (2020). Our approach, however, is frequentist, as are most of the applications in the literature. In particular, we show how to incorporate historical estimates in scenarios which we classify at Type {\rm I} Problems, where the historical parameters are not reestimated, and Type {\rm II} Problems, where they are. Two variants of Type {\rm II} problems are described. See Theorems \ref{Thm.Limit.4.Estimator.I}, \ref{Thm.Limit.4.Estimator.IIa} and \ref{Thm.Limit.4.Estimator.IIb}. We also show that if, given the data $\mathcal{D}$, it is possible to reestimate the historical parameters then it is beneficial to do so at least for large sample sizes (Theorem \ref{Thm.Order.A&B}). Other preference relations, in fact a hierarchy, among the estimators and any function thereof, were also established, cf. Theorems \ref{Thm.Order.C&U}, \ref{Thm.Order.B&C}. It was also demonstrated that the availability of historical data should be taken into account when an optimal experiment is designed. In particular, relevant methods for a two--way ANOVA and for testing drug interaction were discussed. 

In our analysis we have assumed that the the data $\mathcal{D}$ is a random sample and that the estimating equation \eqref{Eq.Type.I} is of an additive form. These assumptions have been used merely to simplify the exposition and are easily modified to dependent data and various other estimating functions. It is clear that Type {\rm I} and {\rm II} Problems describe a broad range of possibilities, nevertheless they are insufficient for describing the rich collection of problems in which historical estimates may play a role. For example, our formulation assumes that the historical parameters $\bm{\eta}_1,\ldots,\bm{\eta}_K$ are distinct. However, in many situations this is not so. In fact, some of the historical studies may be full or partial replicates of each other. In cases when the current study is a partial replicate of a historical study, simple plug-in methods or reestimation methods may be used. One has to be careful, though, about the choice of the estimates. We are aware of situations where a simple plug--in estimator performs better than a less than optimal reestimating method. Throughout, we have assumed interchangeability. Clearly there are many experimental settings, especially in the sciences, where this assumption is realistic. In other situations, say clinical trials, heterogeneity rather than interchangeability is the rule. In such cases some modification of the methods proposed, using random effect models, may be possible. See Rukhin (2007) and the references therein. 

Finally, it is also worth mentioning that the problem of accounting for historical estimates is naturally related, for obvious reasons, to sequential analysis, where data is collected over time, to meta--analysis, where the effort is to combine information from different sources and double sampling, and especially non--nested double sampling (Hidiroglou, 2001), which attempts to provide better inferences by augmenting and predicting unobserved quantities from existing data sets. The literature on combining surveys (Kim and Rao, 2012) is also relevant. Further understanding can be possibly attained by incorporating ideas from these fields.

\section*{Acknowledgments}

Both authors thank COST Action IC1408 for Computationally Intensive Methods for the Robust Analysis of Non--standard Data that supported this research with grants for short visits. In addition the work of Ori Davidov was partially supported by the Israel Science Foundation Grant No. 456/17 and gratefully acknowledged. The authors are indebted to Anna Klimova for drawing their attention to the importance of Bliss independence.

\newpage

\newpage
\begin{center}
\Large{\bf{Tables}}
\end{center}

\begin{table}[h] 
{\centering
\begin{tabular}{c|c|c||c|c||c|c||c|c||c|c||c|c}
$m$& \multicolumn{2}{|c||}{$100$} & \multicolumn{2}{|c||}{$200$} & 
\multicolumn{2}{|c||}{$500$} & \multicolumn{2}{|c||}{$1000$} & 
\multicolumn{2}{|c||}{$2000$} & \multicolumn{2}{|c}{$5000$} \\ \hline
$n$ & $A_{\theta \theta }$ & $B_{\theta \theta }$ & $A_{\theta
\theta }$ & $B_{\theta \theta }$ & $A_{\theta \theta
}$ & $B_{\theta \theta }$ & $A_{\theta \theta }$ & 
$B_{\theta \theta }$ & $A_{\theta \theta }$ & 
$B_{\theta \theta }$ & $A_{\theta \theta }$ & 
$B_{\theta \theta }$ \\ \hline
$100$ & 11.0 & 9.3 &  6.0  & 7.5 & 3.0 & 5.7 & 2.0 & 4.9 & 1.5 & 4.5 & 1.2
& 4.2 \\ 
$200$ & 21.0 & 11.33 &  11.0 & 9.33 &  5.0 & 6.95 &  3.0 & 5.70 &  2.0 & 
4.92 & 1.4 & 4.39 \\ 
$500$ & 51.0 & 13.52 & 26.0 & 11.94 & 11.0 & 9.33 & 6.0 & 7.47 & 3.5 & 6.04
& 2.0 & 4.39 \\ 
$1000$ & 101.0 & 14.61 & 51.0 & 13.52 & 21.0 & 11.3 & 11.0 & 9.33 & 6.0 & 
7.47 & 3.0 & 5.70 \\ 
$2000$ & 201.0 & 15.26 & 101.0 & 14.61 & 41.0 & 13.07 & 21.0 & 11.33 & 11.0
& 9.33 & 5.0 & 6.95 \\ 
$5000$ & 501.0 & 15.69 & 251.0 & 15.4 & 101.0 & 14.61 & 51.0 & 13.52 & 26.0
& 11.94 & 11.0 & 9.33 \\ 
\end{tabular}
\caption{Asymptotic variances for $\theta$ for Type \rm{I} and Type \rm{II} Problems (with a balanced design) as function of the sizes of the the historical and current studies}
\label{Tab.A.B}
}
\end{table}

\begin{table}[h] 
{\centering
\begin{tabular}{c|c|c}
Sampling Ratio $( \rho )$ & Minimal Variance $B_{\theta\theta}$ & Optimal Design $(\boldsymbol{\xi}) $ \\ \hline
$1/8$ & 2.27  & $(0.020,0.001,0.001,0.978)$ \\ 
$1/4$ & 3.51  & $(0.020,0.001,0.001,0.978)$ \\ 
$1/2$ & 5.53  & $(0.020,0.160,0.160,0.660)$ \\ 
$1$   & 8.00  & $(0.020,0.243,0.243,0.494)$ \\ 
$2$   & 10.66 & $(0.125,0.250,0.250,0.375)$  \\ 
$4$   & 12.80 & $(0.187,0.250,0.250,0.313)$ \\ 
$8$   & 14.22 & $(0.219,0.250,0.250,0.281) $ \\ 
\end{tabular}
\caption{Optimal design for Type \rm{II} problems as a function of the sampling ratio $\rho$}
\label{Tab.Opt}
}
\end{table}

\begin{table}[h]
\begin{center}
		\begin{tabular}{ccccccccccccccccc}
			\hline\\
			   $m_1$ & $m_2$ & $n_{min}$ & $n_{12}$ & $n_1$ & $n_2$  & & $m_1$ & $m_2$ & $n_{min}$ & $n_{12}$ & $n_1$ & $n_2$\\
			\hline\\
			\multicolumn{6}{c}{$(\eta_1, \eta_2) = (0.3, 0.3)$} &  & \multicolumn{6}{c}{$(\eta_1, \eta_2) = (0.5, 0.7)$}\\
			\cline{1-6}  \cline{8-13}\\
			     10 & 10 & 23 & 22 & 1 & 0 &  &  10 & 10 & 15 & 14 & 1 & 0\\
			    20 & 10 & 23 & 22 & 0 & 1 & &  10 &20 & 15 & 14 & 1 & 0\\
			   30 & 10 & 23 & 22 & 0 & 1 & &  10 & 30 & 15 & 14 & 1 & 0\\
			   & & & &  &  &  & 20 & 10 & 23 & 22 & 0 & 1\\
			   \multicolumn{6}{c}{$(\eta_1, \eta_2) = (0.3, 0.5)$} & &
			  30 & 10 & 23 & 22 & 0 & 1\\
			  \cline{1-6}  \\
			     10 & 10 & 17 & 16 & 1 & 0 &  & \multicolumn{6}{c}{$(\eta_1, \eta_2) = (0.5, 0.9)$}\\
			     \cline{8-13}
			  20 & 10 & 26 & 25 & 0 & 1 & &\\
			    30 & 10 & 27 & 26 & 0 & 1 & &  10 & 10 & 13 & 12 & 1 & 0\\
			    10 & 20 & 17 & 16 & 1 & 0 & & 10 & 20 & 13 & 12 & 1 & 0\\
			   10 & 30 & 17 & 16 & 1 & 0 & & 10 & 30 & 13 & 12 & 1 & 0\\
			   &  & & & & & & 20 & 10 & 24 & 23 & 1 & 0\\
			  \multicolumn{6}{c}{$(\eta_1, \eta_2) = (0.3, 0.7)$}  & & 30 & 10 & 35 & 34 & 1 & 0\\
			  \cline{1-6}  \\
			   10 & 10 & 14 & 13 & 1 & 0 & & \multicolumn{6}{c}{$(\eta_1, \eta_2) = (0.7, 0.7)$}\\
			  
			  \cline{8-13}
			     10 & 20 & 14 & 13 & 1 & 0 & & \\
			     10 & 30 & 14 & 13 & 1 & 0 & & 10 & 10 & 17 & 16 & 1 & 0\\
			   20 & 10 & 28 & 27 & 1 & 0 & & 20 & 10 & 17 & 16 & 0 & 1\\
			   30 & 10 & 32 & 31 & 0 & 1 & & 30 & 10 & 17 & 16 & 0 & 1\\\\
			   
			   \multicolumn{6}{c}{$(\eta_1, \eta_2) = (0.3, 0.9)$} &  & \multicolumn{6}{c}{$(\eta_1, \eta_2) = (0.7, 0.9)$}\\
			\cline{1-6}  \cline{8-13}\\ 
                10 & 10 & 12 & 11 & 1 & 0& & 10 & 10 & 18 & 17 & 1 & 0\\
              10 & 20 & 12 & 11 & 1 & 0 & & 20 & 10 & 25 & 24 & 1 & 0\\
               10 & 30 & 12 & 11 & 1 & 0 &  & 30 & 10 & 25 & 24 & 0 & 1\\
              20 & 10 & 23 & 22 & 1 & 0 & & 10 & 20 & 13 & 12 & 1 & 0\\
              30 & 10 & 34 & 33 & 1 & 0 & & 10 & 30 & 13 & 12 & 1 & 0\\

			 \hline
		\end{tabular}\label{Table2_Example12}
	\end{center}
	\caption{The minimal sample size $n_{min}$, at which optimal allocation requires to repeat historical observations, for selected values of the and no-success probabilities $\eta_1$, $\eta_2$ and historical sample sizes $m_1$, $m_2$ in a test of drug interaction.}
	\label{t1}
\end{table}

\newpage
\newpage

\begin{center}
\Large{\bf{Appendix: Proofs}}
\end{center}

\section*{{Proof of Theorem \ref{Thm.Limit.4.Estimator.I}:}}

\begin{proof}
Since $\bar{\bm{\theta}}_{A}$ solves \eqref{Eq.Type.I} we have
\begin{equation} \label{Eq1.Th1}
\bm{\Psi}(\bar{\bm{\theta}}_{A},\widehat{\bm{\eta}})=\bm{0}.    
\end{equation}
By assumption $\bm{\psi}$ is continuous and differentiable with respect to $\bm{\theta}$ and $\bm{\eta}_{1},\ldots,\bm{\eta}_K$. Thus, so is $\bm{\Psi}$. Hence, by the mean value theorem
\begin{equation*}
\bm{\Psi}(\bar{\bm{\theta}}_{A},\widehat{\bm{\eta}})=\bm{\Psi}(\bm{\theta}_{0},\widehat{\bm{\eta }})+\frac{\partial }{\partial\bm{\theta}}{\bm\Psi}(\bm{\theta}_{0},\widehat{\bm{\eta}})(\bar{\bm{\theta}}_{A}-\bm{\theta}_{0})+o(||\bar{\bm{\theta}}_{A}-\bm{\theta}_{0}||).    
\end{equation*}
Applying the mean value theorem to $\bm{\Psi}(\bm{\theta}_{0},\widehat{\bm{\eta}})$ in the display above yields
\begin{equation*}
\bm{\Psi}(\bm{\theta}_{0},\widehat{\bm{\eta}})= \bm{\Psi}(\bm{\theta}_{0},\bm{\eta}_{0})+\frac{\partial }{\partial \bm{\eta}}\bm{\Psi}(\bm{\theta}_{0},{\bm{\eta}}_{0})(\widehat{\bm{\eta}}_-\bm{\eta}_{0}) + o(||\widehat{\bm{\eta}}-\bm{\eta}_{0}||), 
\end{equation*}
so \eqref{Eq1.Th1} can be rewritten as 
\begin{equation*}
\bm{\Psi}(\bm{\theta}_{0},\bm{\eta}_{0})+\frac{\partial }{\partial \bm{\eta}}\bm{\Psi}(\bm{\theta}_{0},{\bm{\eta}}_{0})(\widehat{\bm{\eta}}-\bm{\eta}_{0})+\frac{\partial }{\partial\bm{\theta}}{\bm\Psi}(\bm{\theta}_{0},\widehat{\bm{\eta}})(\bar{\bm{\theta}}_{A}-\bm{\theta}_{0})+\bm{R}=0
\end{equation*}
where, assuming consistency $\bm{R}=o(||\bar{\bm{\theta}}_{A}-\bm{\theta}_{0}||)+o(||\widehat{\bm{\eta}}-\bm{\eta}_{0}||)=o_{p}(1)$. Now, by the continuous mapping theorem and the law of large numbers we have:
\begin{eqnarray*}
\frac{\partial}{\partial \bm{\theta}}\bm{\Psi}(\bm{\theta}_0,\widehat{\bm{\eta}}) 
&=&\frac{1}{n}\sum_{i=1}^{n}\frac{\partial }{\partial 
\bm{\theta }}\bm{\psi}(\bm{\theta}_0,\widehat{\bm{\eta}},\bm{Y}_{i})\\
&=&\frac{1}{n}\sum_{i=1}^{n}\frac{\partial }{\partial \bm{\theta }}\bm{\psi}(\bm{\theta }_{0},\bm{\eta }_{0},\bm{Y}_{i})+o_{p}(1) = \mathbb{E}_{0}(\partial \bm{\psi}/\partial\bm{\theta}) +o_{p}(1)  
\end{eqnarray*}
which is a $p\times p$ matrix. Similarly,
\begin{eqnarray*}
\frac{\partial}{\partial \bm{\eta}}\bm{\Psi} (\bm{\theta}_{0},{\bm{\eta}}_{0}) 
&=&\frac{1}{n}\sum_{i=1}^{n}\frac{\partial }{\partial 
\bm{\eta }}\bm{\psi}(\bm{\theta}_{0},\bm{\eta}_{0},
\bm{Y}_{i}) = \mathbb{E}_{0}(\partial \bm{\psi}
/\partial \bm{\eta })+o_{p}(1),  
\end{eqnarray*}
which is a $p\times q$ matrix. For convenience we set $\bm{D}_{\bm{\theta}_0}=\mathbb{E}_{0}(\partial \bm{\psi}/\partial \bm{\theta})$ and $\bm{D}_{\bm{\eta}_0}=\mathbb{E}_{0}(\partial\bm{\psi}
/\partial\bm{\eta})$. Hence we can reexpress  \eqref{Eq1.Th1} more concisely as
\begin{equation*}
\bm{\Psi}(\bm{\theta }_{0},\bm{\eta }_{0})+\bm{D}_{\bm{\eta}_0}
(\widehat{\bm{\eta }}-\bm{\eta}_{0})+\bm{D}_{\bm{\theta}_{0}}(\bar{\bm{\theta }}_{A}-\bm{\theta }_{0})+o_{p}\left( 1\right) =0
\end{equation*}
from which it follows, by the invertability of $\bm{D}_{\bm{\theta}_0}$, that
\begin{equation} \label{Eq.2.Pf.I}
\sqrt{n}(\bar{\bm{\theta}}_{A}-\bm{\theta }_{0})=-
\bm{D}_{\bm{\theta}_{0}}^{-1}\{\frac{1}{\sqrt{n}}\sum_{i=1}^{n}\bm{\psi}(\bm{\theta }_{0},\bm{\eta}_{0},\bm{Y}_{i})+\frac{\sqrt{n}}{\sqrt{m}} \bm{D}_{\bm{\eta}_0}\sqrt{m}(\widehat{\bm{
\eta }}-\bm{\eta }_{0})\}+o_{p}\left( 1\right) .
\end{equation}
Since the first term in the curly brackets above is a function of the data $\mathcal{D}$ collected in $\mathcal{S}$ and the second term depends on the historical data, i.e., the studies $\mathcal{S}_1,\ldots,\mathcal{S}_K$ the two terms are independent. Now, by the central limit theorem
\begin{equation*}
\frac{1}{\sqrt{n}}\sum_{j=1}^{n}\bm{\psi}(\bm{\theta}_{0},\bm{\eta}_{0},\bm{Y}_i) \Rightarrow \mathcal{N}_{p}(\bm{0},\bm{\Sigma}_{\bm{\psi}})
\end{equation*}
where $\bm{\Sigma}_{\bm{\psi}}=\mathbb{E}_{0}(\bm{\psi\psi}^{T}).$ By assumption $\sqrt{m_{j}}(
\widehat{\bm{\eta }}_{j}-\bm{\eta }_{j,0})\Rightarrow 
\mathcal{N}_{q_{j}}(\bm{0},\bm{\Sigma }_{j})$ for each $j$. Thus $\sqrt{m}(\widehat{\bm{\eta}}-\bm{\eta}_{0}) \Rightarrow 
\mathcal{N}_{q}(\bm{0},\bm{\Sigma})$ where $\bm{\Sigma} = \mathrm{BlockDiag}(\kappa_{1}\bm{\Sigma }_{1},\ldots,\kappa_{K}\bm{\Sigma}_{K})$ with $\kappa_{j} = \lim(m/m_j)$ for $j=1,\ldots,K$. Thus,
\begin{equation}
\frac{\sqrt{n}}{\sqrt{m}} \bm{D}_{\bm{\eta}_0}\sqrt{m}(\widehat{\bm{\eta }}-\bm{\eta }_{0}) \Rightarrow \mathcal{N}_{p}(\bm{0},\rho\bm{D}_{\bm{\eta}_{0}}\bm{\Sigma }\bm{D}_{\bm{\eta}_{0}}^{T}).   
\end{equation}
Collecting terms shows that $\sqrt{n}(\bar{\bm{\theta}}_{A}-\bm{\theta }_{0})\Rightarrow 
\mathcal{N}_{p}(\bm{0},\bm{A}_{\bm{\theta\theta}})$ where $\bm{A}_{\bm{\theta\theta}}$ is as stated. Now, recall that $\bar{\bm{\eta}}_{A}=\widehat{\bm{\eta}}$. Thus, marginally $\sqrt{n}(\bar{\bm{\eta}}_{A}-\bm{\eta}_{0})\Rightarrow \mathcal{N}_{q}(\bm{0},\rho\bm{\Sigma})$, so $\bm{A}_{\bm{\eta\eta}} = \rho\bm{\Sigma}$. Clearly the joint asymptotic distribution of $\sqrt{n}(\bar{\bm{\theta}}_{A}-\bm{\theta}_{0},\bar{\bm{\eta}}_{A}-\bm{\eta }_{0})^{T}$, is also multivariate normal. Thus, 
\begin{equation*}
\bm{A}_{\bm{\theta\eta}} =\lim_{n}{\rm Cov}(\sqrt{n}(\bar{\bm{\theta}}_{A}-\bm{\theta}_{0}),\sqrt{n}(\bar{\bm{\eta}}_{A}-\bm{\eta }_{0})) = \lim_{n}  [-\bm{D}_{\bm{\theta}_{0}}^{-1}\bm{D}_{\bm{\eta}_{0}}{\rm Cov}(\sqrt{n}(\bar{\bm{\eta }}_{A}-\bm{\eta }_{0}),\sqrt{n}(\bar{\bm{\eta }}_{A}-\bm{\eta }_{0}))]= -\rho\bm{D}_{\bm{\theta}_{0}}^{-1}\bm{D}_{\bm{\eta}_{0}}\bm{\Sigma}, \end{equation*}
as required, completing the proof.
\end{proof}

\section*{{Proof of Theorem \ref{Thm.Limit.4.Estimator.IIa}:}}

\begin{proof}
Since $\bar{\bm{\theta}}_{B}$ solves \eqref{Eq.Type.II.Step.2} where $\bar{\bm{\eta}}$ is given in \eqref{Eq.eta.bar} we have
\begin{equation} \label{Eq.I.Pf.II}
\bm{\Psi}(\bar{\bm{\theta}}_{B},\bar{\bm{\eta}})=\bm{0}.    
\end{equation}
Following the derivations in the proof of Theorem \ref{Thm.Limit.4.Estimator.I}, but with $\bar{\bm{\eta}}$ instead of $\widehat{\bm{\eta}}$, we find that \eqref{Eq.I.Pf.II} can be rewritten as 
\begin{equation*}
\bm{\Psi}(\bm{\theta }_{0},\bm{\eta }_{0})+\bm{D}_{\bm{\eta}_0}
(\bar{\bm{\eta }}-\bm{\eta}_{0})+\bm{D}_{\bm{\theta}_{0}}(\bar{\bm{\theta}}_{B}-\bm{\theta }_{0})+o_{p}\left( 1\right) =\bm{0}
\end{equation*}
from which it follows, by the invertibility of $\bm{D}_{\bm{\theta}_0}$, that
\begin{equation} \label{Eq.II.Pf.II}
\sqrt{n}(\bar{\bm{\theta}}_{B}-\bm{\theta }_{0})=-
\bm{D}_{\bm{\theta}_{0}}^{-1}\{\frac{1}{\sqrt{n}}\sum_{j=1}^{n}\bm{\psi}(\bm{\theta }_{0},\bm{\eta}_{0},\bm{Y}_{i})+ \bm{D}_{\bm{\eta}_0}\sqrt{n}(\bar{\bm{
\eta }}-\bm{\eta }_{0})\}+o_{p}\left( 1\right) .
\end{equation}
Now using \eqref{Eq.eta.bar} and \eqref{Eq.W1&W2} we find that
\begin{equation*}
\sqrt{n}(\bar{\bm{\eta }}-\bm{\eta}_{0}) =  \sqrt{n}(\bm{W}_{1}(\tilde{\bm{\eta}}-\bm{\eta}_{0})+ \bm{W}_{2}(\widehat{\bm{\eta}}-\bm{\eta}_{0}))+o_{p}(1)  
\end{equation*}
which we may substitute into \eqref{Eq.II.Pf.II} yielding
\begin{equation} \label{Eq.III.Pf.II}
\sqrt{n}(\bar{\bm{\theta}}_{B}-\bm{\theta }_{0})=-
\bm{D}_{\bm{\theta}_{0}}^{-1}\{\frac{1}{\sqrt{n}}\sum_{j=1}^{n}\bm{\psi}(\bm{\theta }_{0},\bm{\eta}_{0},\bm{Y}_{i})+ \bm{D}_{\bm{\eta}_0}\bm{W}_1\sqrt{n}(\tilde{\bm{\eta}}-\bm{\eta}_{0})
+\bm{D}_{\bm{\eta}_0}\bm{W}_2\sqrt{\frac{n}{m}} \sqrt{m}(\widehat{\bm{\eta}}-\bm{\eta}_{0})
\}+o_{p}\left( 1\right) .
\end{equation}
The three terms in the curly brackets in \eqref{Eq.III.Pf.II} satisfy:
\begin{align*}
\frac{1}{\sqrt{n}}\sum_{i=1}^{n}\bm{\psi}(\bm{\theta}_{0},\bm{\eta}_{0},\bm{Y}_i)  & \Rightarrow \mathcal{N}_{p}(\bm{0},\bm{\Sigma}_{\bm{\psi}}), \\
\bm{D}_{\bm{\eta}_0}\bm{W}_1\sqrt{n}(\tilde{\bm{\eta }}-\bm{\eta }_{0}) & \Rightarrow \mathcal{N}_{q}(\bm{0},\bm{D}_{\bm{\eta}_0}\bm{W}_1\bm{\Upsilon}_{\bm{\eta\eta}}\bm{W}_1^{T}\bm{D}_{\bm{\eta}_0}^{T}), \\
\bm{D}_{\bm{\eta}_0}\bm{W}_2\frac{\sqrt{n}}{\sqrt{m}}\sqrt{m}(\widehat{\bm{\eta }}-\bm{\eta }_{0}) & \Rightarrow \mathcal{N}_{q}(\bm{0},\rho\bm{D}_{\bm{\eta}_{0}}\bm{W}_2\bm{\Sigma }\bm{W}_2^{T}\bm{D}_{\bm{\eta}_{0}}^{T}).
\end{align*}
Now, since $\tilde{\bm{\eta}}$ is a solution to \eqref{Eq.Type.II.Step.1} the quantity $\sqrt{n}(\tilde{\bm{\eta }}-\bm{\eta }_{0})$ may be expressed as 
$$ 
\frac{1}{\sqrt{n}} \sum_{i=1}^{n}\bm{\varphi}(\bm{\theta}_{0},\bm{\eta}_{0},\bm{Y}_i)+o_p(1)
$$ 
for some function $\bm{\varphi}$ which is known as the influence function (cf. Van der Vaart, 2000). For more details see Remark \ref{Rem.More.B&C}. It follows by 
the central limit theorem, that 
\begin{equation*}
\frac{1}{\sqrt{n}} \sum_{i=1}^{n}\bm{\psi}(\bm{\theta}_{0},\bm{\eta}_{0},\bm{Y}_i) \quad \makebox{and} \quad   \frac{1}{\sqrt{n}} \sum_{i=1}^{n}\bm{\varphi}(\bm{\theta}_{0},\bm{\eta}_{0},\bm{Y}_i)
\end{equation*}
are jointly multivariate normal, thus so are the first two terms in the curly brackets of \eqref{Eq.III.Pf.II}. Moreover the third term, which depends on the historical data is independent of the first two terms and normally distributed. Now the covariance among the first two terms is 
\begin{align*}
{\rm Cov}(\frac{1}{\sqrt{n}}\sum_{i=1}^{n}\bm{\psi}(\bm{\theta}_{0},\bm{\eta}_{0},\bm{Y}_i),\sqrt{n}(\tilde{\bm{\eta }}-\bm{\eta }_{0})) &= {\rm Cov}(\sum_{i=1}^{n}\bm{\psi}(\bm{\theta}_{0},\bm{\eta}_{0},\bm{Y}_i),(\tilde{\bm{\eta }}-\bm{\eta }_{0})) = {\rm Cov}(n\bm{\psi}(\bm{\theta}_{0},\bm{\eta}_{0},\bm{Y}_1),(\tilde{\bm{\eta }}-\bm{\eta }_{0})) \\
& = \mathbb{E}_{0}(\bm{\psi}(\bm{\theta}_{0},\bm{\eta}_{0},\bm{Y}_1) n(\tilde{\bm{\eta }}-\bm{\eta }_{0})) = \mathbb{E}_{0}(\mathbb{E}_{0}(\bm{\psi}(\bm{\theta}_{0},\bm{\eta}_{0},\bm{Y}_1) n(\tilde{\bm{\eta }}-\bm{\eta }_{0})|\bm{Y}_1)) \\
& = \mathbb{E}_{0}(\bm{\psi}(\bm{\theta}_{0},\bm{\eta}_{0},\bm{Y}_1)\mathbb{E}_{0}(n(\tilde{\bm{\eta }}-\bm{\eta }_{0})|\bm{Y}_1)).
\end{align*}
However, by assumption $\tilde{\bm{\eta}}$ is asymptotically unbiased and $\sqrt{n}$--consistent, i.e., $\mathbb{E}_{0}(\tilde{\bm{\eta}}) = \bm{\eta}_0 + b/n + o(1/n)$ so $\mathbb{E}_{0}(n(\tilde{\bm{\eta }}-\bm{\eta }_{0})|\bm{Y}_1))=n\mathbb{E}_{0}((\tilde{\bm{\eta }}-\bm{\eta }_{0}))+o(1) = O(1)$. Plugging the latter into the above display shows that covariance above converges to $0$ as $n\rightarrow\infty$. It now follows that all three terms appearing in \eqref{Eq.III.Pf.II} are asymptotically independent. 

Set $\bar{\bm{\eta }}_{B}=\bar{\bm{\eta }}$ and observe that using  \eqref{Eq.III.Pf.II} we have $\sqrt{n}(\bar{\bm{\eta}}_{B}-\bm{\eta}_{0})\Rightarrow \mathcal{N}_{q}(\bm{0},\bm{B}_{\bm{\eta\eta}})$ where $$\bm{B}_{\bm{\eta\eta}}=\bm{W}_1\bm{\Upsilon}_{\bm{\eta\eta}} \bm{W}_1^{T} + \rho \bm{W}_2\bm{\Sigma}\bm{W}_2^{T}.$$ 
Since $\gamma/(1-\gamma) =\rho$ we may reexpress the weight matrices as $\bm{W}_1=(\bm{\Upsilon}_{\bm{\eta\eta}}^{-1}+(\rho\bm{\Sigma})^{-1})^{-1}\bm{\Upsilon}_{\bm{\eta\eta}}^{-1}$ and $\bm{W}_2=(\bm{\Upsilon}_{\bm{\eta\eta}}^{-1}+(\rho\bm{\Sigma})^{-1})^{-1}(\rho\bm{\Sigma})^{-1}$. Now, using the fact that products of symmetric matrices commute and a bit of algebra it can be shown that
\begin{equation*}
\bm{B}_{\bm{\eta\eta}} = (\bm{\Upsilon}_{\bm{\eta\eta}}^{-1}+(\rho\bm{\Sigma})^{-1})^{-1}.    
\end{equation*}
Collecting terms shows that $\sqrt{n}(\bar{\bm{\theta}}_{B}-\bm{\theta }_{0})\Rightarrow 
\mathcal{N}_{p}(\bm{0},\bm{B}_{\bm{\theta\theta}})$ where $\bm{B}_{\bm{\theta\theta}}$ is as stated. The stochastic representation \eqref{Eq.2.Pf.I} shows that the joint asymptotic distribution of $\sqrt{n}(\bar{\bm{\theta}}-\bm{\theta}_{0},\bar{\bm{\eta }}-\bm{\eta }_{0})$ is also multivariate normal with
\begin{equation*}
{\rm Cov}(\sqrt{n}(\bar{\bm{\theta}}_{B}-\bm{\theta}_{0}),\sqrt{n}(\bar{\bm{\eta }}_{B}-\bm{\eta}_{0})) =  -\bm{D}_{\bm{\theta}_{0}}^{-1}\bm{D}_{\bm{\eta}_{0}}{\rm Cov}(\sqrt{n}(\bar{\bm{\eta }}_{B}-\bm{\eta }_{0}),\sqrt{n}(\bar{\bm{\eta }}_{B}-\bm{\eta }_{0}))\rightarrow -\bm{D}_{\bm{\theta}_{0}}^{-1}\bm{D}_{\bm{\eta}_{0}}\bm{B}_{\bm{\eta\eta}}
\end{equation*}
as required, completing the proof.
\end{proof}

\section*{{Proof of Theorem \ref{Thm.Order.A&B}:}}

The following preliminary Lemma will be used.

\begin{lemma} \label{Lemma.Loewener}
Let $\bm{X}_{1}$ and $\bm{X}_{2}$ be random vectors with variances $\bm{V}_{1}=\mathbb{V}(\bm{X}_{1})$ and $\bm{V}_{2}=\mathbb{V}(\bm{X}_{2})$ with $\bm{V}_{1} \preceq \bm{V}_{2}$. Then for any matrix $\bm{A}$ we have $\mathbb{V}(\bm{A}\bm{X}_{1}) \preceq \mathbb{V}(\bm{A}\bm{X}_{2})$. As a consequence we also have $\bm{V}_{1}^{-1} \succeq \bm{V}_{2}^{-1} $. 
\end{lemma}

\section*{{Proof of Lemma \ref{Lemma.Loewener}:}}

\begin{proof}
Observe that
\begin{align*}
\bm{u}^{T}\mathbb{V}(\bm{A}\bm{X}_{1})\bm{u} &=  \bm{u}^{T}\bm{A}\bm{V}_{1}\bm{A}^{T}\bm{u} = (\bm{A}^{T}\bm{u})^{T}\bm{V}_{1}(\bm{A}^{T}\bm{u}) = \bm{v}^{T}\bm{V}_{1}\bm{v} \preceq \bm{v}^{T}\bm{V}_{2}\bm{v} = (\bm{A}^{T}\bm{u})^{T}\bm{V}_{2}(\bm{A}^{T}\bm{u}) \\ 
&= \bm{u}^{T}\bm{A}\bm{V}_{2}\bm{A}^{T}\bm{u} = \bm{u}^{T}\mathbb{V}(\bm{A}\bm{X}_{2})\bm{u}.
\end{align*}
for any vector $\bm{u}$. The inequality $\bm{v}^{T}\bm{V}_{1}\bm{v} \preceq \bm{v}^{T}\bm{V}_{2}\bm{v}$ holds since $\bm{V}_2-\bm{V}_1$ is non--negative definite by assumption. Thus $\mathbb{V}(\bm{A}\bm{X}_{1}) \preceq \mathbb{V}(\bm{A}\bm{X}_{2})$ as claimed. 

Now choose $\bm{A}=\bm{V}_{1}^{-1/2}\bm{V}_{2}^{-1/2}$ and note that
\begin{align*}
\mathbb{V}(\bm{A}\bm{X}_{1}) &=  (\bm{V}_{1}^{-1/2}\bm{V}_{2}^{-1/2}) \bm{V}_{1} (\bm{V}_{1}^{-1/2}\bm{V}_{2}^{-1/2})^{T} = \bm{V}_{2}^{-1} \\  
\mathbb{V}(\bm{A}\bm{X}_{2}) &=  (\bm{V}_{1}^{-1/2}\bm{V}_{2}^{-1/2}) \bm{V}_{2} (\bm{V}_{1}^{-1/2}\bm{V}_{2}^{-1/2})^{T} = \bm{V}_{1}^{-1}.
\end{align*}
The equalities above hold since products of symmetric matrices commute. The inequality $\bm{V}_{1}^{-1} \succeq \bm{V}_{2}^{-1} $ follows immediately.
\end{proof}

We now continue with the proof of Theorem \ref{Thm.Order.A&B}:

\begin{proof}
Observe that $\bm{A}$ is the variance matrix of the random vector
\begin{equation*}
\bm{T}_1 =  \begin{pmatrix}
-\bm{D}_{\bm{\theta}_0}^{-1}  & -\bm{D}_{\bm{\theta}_0}^{-1}\bm{D}_{\bm{\eta}_0} \\
\bm{0}   & \bm{I}
\end{pmatrix}
\begin{pmatrix}
\bm{S}_1 \\ \bm{S}_2
\end{pmatrix}
\end{equation*}
where $\bm{S}_1 \sim \mathcal{N}_{p}(\bm{0},\bm{\Sigma}_{\bm{\psi}})$ and $\bm{S}_2 \sim \mathcal{N}_{q}(\bm{0},\rho\bm{\Sigma})$ are independent. Similarly, $\bm{B}$ is the variance matrix of the random vector
\begin{equation*}
\bm{T}_2 =  \begin{pmatrix}
-\bm{D}_{\bm{\theta}_0}^{-1}  & -\bm{D}_{\bm{\theta}_0}^{-1}\bm{D}_{\bm{\eta}_0} \\
\bm{0}   & \bm{I}
\end{pmatrix}
\begin{pmatrix}
\bm{S}_1 \\ \bm{S}_3
\end{pmatrix}
\end{equation*}
where $\bm{S}_1 \sim \mathcal{N}_{p}(\bm{0},\bm{\Sigma}_{\bm{\psi}})$ and $\bm{S}_3 \sim \mathcal{N}_{q}(\bm{0},(\bm{\Upsilon}_{\bm{\eta\eta}}+(\rho\bm{\Sigma})^{-1})^{-1}$ are independent. Now,
\begin{equation*}
\mathbb{V}\begin{pmatrix}
\bm{S}_1 \\ \bm{S}_2
\end{pmatrix}-\mathbb{V} \begin{pmatrix}
\bm{S}_1 \\ \bm{S}_3
\end{pmatrix} = 
\begin{pmatrix}
\bm{0} & \bm{0} \\
\bm{0} & \rho\bm{\Sigma}-(\bm{\Upsilon}_{\bm{\eta\eta}}+(\rho\bm{\Sigma})^{-1})^{-1}
\end{pmatrix}
\end{equation*}
It is easy to verify that $\bm{\Upsilon}_{\bm{\eta\eta}}+(\rho\bm{\Sigma})^{-1} \succeq (\rho\bm{\Sigma})^{-1}$ so by the second part of Lemma \ref{Lemma.Loewener} we have $\rho\bm{\Sigma} \succeq (\bm{\Upsilon}_{\bm{\eta\eta}}+(\rho\bm{\Sigma})^{-1})^{-1}$ and therefore
\begin{equation*}
\mathbb{V} \begin{pmatrix}
\bm{S}_1 \\ \bm{S}_3
\end{pmatrix} \preceq
\mathbb{V} \begin{pmatrix}
\bm{S}_1 \\ \bm{S}_2
\end{pmatrix}.
\end{equation*}
Applying Lemma \ref{Lemma.Loewener} we find that
\begin{equation*}
\bm{B}=\mathbb{V}(\bm{T}_2) \preceq   \mathbb{V}(\bm{T}_1) = \bm{A} 
\end{equation*}
as required.  

Next, an application of the $\delta$--method and Theorems \ref{Thm.Limit.4.Estimator.I} and \ref{Thm.Limit.4.Estimator.IIa} shows that $\sqrt{n}(\bm{\Phi}(\bar{\bm{\theta}}_A,\bar{\bm{\eta}}_A)-\bm{\Phi}(\bm{\theta}_0,\bm{\eta}_0)) \Rightarrow \mathcal{N}_{r}(\bm{0},\bm{P}\bm{A}\bm{P}^{T})$ and $\sqrt{n}(\bm{\Phi}(\bar{\bm{\theta}}_B,\bar{\bm{\eta}}_B)-\bm{\Phi}(\bm{\theta}_0,\bm{\eta}_0)) \Rightarrow \mathcal{N}_{r}(\bm{0},\bm{P}\bm{B}\bm{P}^{T})$    
where $r$ is the dimension of $\bm{\Phi}$ and $\bm{P}=\mathbb{E}_0(\partial\Phi/\partial{\bm{\omega}})$. Observe that $\bm{P}\bm{A}\bm{P}^{T}$ is the variance of the random vector $\bm{P}\bm{T}_1$ whereas $\bm{P}\bm{B}\bm{P}^{T}$ is the variance $\bm{P}\bm{T}_2$. Since $\bm{B} \preceq \bm{A}$ it follows from Lemma \ref{Lemma.Loewener} that $\bm{P}\bm{B}\bm{P}^{T} \preceq \bm{P}\bm{A}\bm{P}^{T}$ concluding the proof. 
\end{proof}

The following lemma motivates the use of the estimators \eqref{Eq.eta.bar} and \eqref{Eq.theta.bar}

\begin{lemma} \label{Lemma.Estimators.II}
Let $\bm{W}\sim\mathcal{N}_{q}(\bm{\eta},m^{-1}\bm{\Sigma})$ and $(\bm{U},\bm{V})^{T}\sim\mathcal{N}_{p+q}((\bm{\theta},\bm{\eta })^{T},n^{-1}\bm{\Upsilon})$ be independent random vectors where 
\begin{equation*}
\bm{\Upsilon }=\left( 
\begin{array}{cc}
\bm{\Upsilon }_{\bm{\theta\theta}} & \bm{\Upsilon }_{{\bm{\theta
\eta}}} \\ 
\bm{\Upsilon}_{\bm{\eta\theta}} & \bm{\Upsilon }_{\bm{\eta\eta} }
\end{array}
\right) .
\end{equation*}
Then the MLEs of $\bm{\theta}$ and $\bm{\eta}$ are
\begin{equation} \label{eq.MLEs.lemma}
\bar{\bm{\theta}} = \bm{U} - \bm{\Upsilon}_{\bm{\theta\eta}}\bm{\Upsilon}_{\bm{\eta\eta}}^{-1}(\bm{V}-  \bar{\bm{\eta}}) \quad \textrm{and} \quad \bar{\bm{\eta}} = (n\bm{\Upsilon}_{\bm{\eta\eta}}^{-1}+m\bm{\Sigma}^{-1}) (n\bm{\Upsilon}_{\bm{\eta\eta}}^{-1}\bm{V}+m\bm{\Sigma}^{-1}\bm{W}).
\end{equation}
\end{lemma}

\section*{{Proof of Lemma \ref{Lemma.Estimators.II}:}}

\begin{proof}
The likelihood is given by
\begin{eqnarray*}
L(\bm{\theta },\bm{\eta }) &=&f(\bm{U},\bm{V};\bm{\theta},\bm{\eta})f(\bm{W};\bm{\eta })
\\
&=&f(\bm{U}|\bm{V};\bm{\theta },\bm{\eta})f(\bm{V};\bm{\eta})f(\bm{W};\bm{\eta}).
\end{eqnarray*}
Now $\bm{U}|\bm{V} \sim \mathcal{N}_{p}(\bm{\lambda},\bm{\Lambda})$ with
\begin{eqnarray*}
\bm{\lambda} &=& \mathbb{E}(\bm{U}|\bm{V})= \bm{\theta} + \bm{\Upsilon }_{\bm{\theta\eta}}\bm{\Upsilon }_{\bm{\eta\eta}}^{-1}(\bm{V}-\bm{\eta}), 
\\
\bm{\Lambda} &=& \mathbb{V}(\bm{U}|\bm{V})= n^{-1}(\bm{\Upsilon }_{\bm{\theta\theta}}-\bm{\Upsilon}_{\bm{\theta\eta}}\bm{\Upsilon}_{\bm{\eta\eta}}^{-1}\bm{\Upsilon}_{\bm{\eta\theta}}),
\end{eqnarray*}
so $\bm{\lambda}$ is linear in both $\bm{\theta}$ and $\bm{\eta}$. Thus we may reparameterize $f(\bm{U}|\bm{V};\bm{\theta},\bm{\eta})$ as $f(\bm{U}|\bm{V};\bm{\lambda})$ where
\begin{equation*}
f(\bm{U}|\bm{V};\bm{\lambda}) \propto \exp\{-\frac{1}{2}(\bm{U}-\bm{\lambda})^T\bm{\Lambda}^{-1} (\bm{U}-\bm{\lambda})\}. 
\end{equation*}
Also marginally $\bm{V}$ follows a $\mathcal{N}_{q}(\bm{\eta},n^{-1}\bm{\Upsilon}_{\bm{\eta\eta}})$ distribution so
\begin{equation*}
f(\bm{V};\bm{\eta})f(\bm{W};\bm{\eta}) \propto \exp\{-\frac{1}{2}(\bm{V}-\bm{\eta})^T n\bm{\Upsilon}_{\bm{\eta\eta}}^{-1} (\bm{V}-\bm{\eta})\}  \exp\{-\frac{1}{2}(\bm{W}-\bm{\eta})^T m\bm{\Sigma}^{-1} (\bm{W}-\bm{\eta})\}   
\end{equation*}
It now follows that the MLEs for $(\bm{\lambda},\bm{\eta})$ are
\begin{eqnarray*}
\bar{\bm{\lambda}} &=& \bm{U}, 
\\
\bar{\bm{\eta}} &=& (n\bm{\Upsilon}_{\bm{\eta\eta}}^{-1}+ m\bm{\Sigma}^{-1})^{-1} (n\bm{\Upsilon}_{\bm{\eta\eta}}^{-1}\bm{V}+ m\bm{\Sigma}^{-1}\bm{U}).
\end{eqnarray*}
Thus by the invariance property of MLEs we find that the MLE of $\bm{\theta}$ is
\begin{equation*}
\bar{\bm{\theta}} = \bm{U} - \bm{\Upsilon}_{\bm{\theta\eta}}\bm{\Upsilon}_{\bm{\eta\eta}}^{-1}(\bm{V}- \bar{\bm{\eta}})    
\end{equation*}
which completes the proof.
\end{proof}

\begin{remark} \label{remark.estimators.II}
To obtain the estimators the estimators \eqref{Eq.eta.bar} and \eqref{Eq.theta.bar} apply Lemma \ref{Lemma.Estimators.II} and substitute $\tilde{\bm{\theta}}$ for $\bm{U}$, $\tilde{\bm{\eta}}$ for $\bm{V}$ and $\widehat{\bm{\eta}}$ for  $\bm{W}$. Further substitute $\tilde{\bm{\Upsilon}}$ and $\widehat{\bm{\Sigma}}$ for $\bm{\Upsilon}$ and $\bm{\Sigma}$, respectively.   
\end{remark}

\section*{{Proof of Theorem \ref{Thm.Limit.4.Estimator.IIb}:}}

\begin{proof}
First note that the difference $\tilde{\bm{\eta}}-\bar{\bm{\eta}}_{C}$ in \eqref{Eq.theta.bar} is a linear combination of $\tilde{\bm{\eta}}$ and $\widehat{\bm{\eta}}$ given by
\begin{equation*}
(n\tilde{\bm{\Upsilon}}_{\bm{\eta\eta}}^{-1}+m\widehat{\bm{\Sigma}}^{-1})^{-1}m\widehat{\bm{\Sigma}}^{-1}(\tilde{\bm{\eta}}-\widehat{\bm{\eta}}).
\end{equation*}
Therefore,
\begin{equation*}
\begin{pmatrix}
\bar{\bm{\theta}}_{C} \\
\bar{\bm{\eta}}_{C}
\end{pmatrix} = 
\begin{pmatrix}
\bm{I}  & -\tilde{\bm{\Upsilon}}_{\bm{\theta\eta}}\tilde{\bm{\Upsilon}}_{\bm{\eta\eta}}^{-1}(n\tilde{\bm{\Upsilon}}_{\bm{\eta\eta}}^{-1}+m\widehat{\bm{\Sigma}}^{-1})^{-1}m\widehat{\bm{\Sigma}}^{-1} & \tilde{\bm{\Upsilon}}_{\bm{\theta\eta}}\tilde{\bm{\Upsilon}}_{\bm{\eta\eta}}^{-1}(n\tilde{\bm{\Upsilon}}_{\bm{\eta\eta}}^{-1}+m\widehat{\bm{\Sigma}}^{-1})^{-1}m\widehat{\bm{\Sigma}}^{-1} \\
\bm{0} & (n\tilde{\bm{\Upsilon}}_{\bm{\eta\eta}}^{-1}+m\widehat{\bm{\Sigma}}^{-1})^{-1}n\tilde{\bm{\Upsilon}}_{\bm{\eta\eta}}^{-1} & (n\tilde{\bm{\Upsilon}}_{\bm{\eta\eta}}^{-1}+m\widehat{\bm{\Sigma}}^{-1})^{-1}m\widehat{\bm{\Sigma}}^{-1}
\end{pmatrix}
\begin{pmatrix}
\tilde{\bm{\theta}} \\
\tilde{\bm{\eta}}   \\
\widehat{\bm{\eta}}
\end{pmatrix}.
\end{equation*}
Since $\bm{\Upsilon}$ and $\bm{\Sigma}$ can be consistently estimated it follows that
\begin{equation*}
\begin{pmatrix}
\bar{\bm{\theta}}_{C} \\
\bar{\bm{\eta}}_{C}
\end{pmatrix} = 
\begin{pmatrix}
\bm{I}  & -\bm{\Upsilon}_{\bm{\theta\eta}}\bm{\Upsilon}_{\bm{\eta\eta}}^{-1}(n\bm{\Upsilon}_{\bm{\eta\eta}}^{-1}+m\bm{\Sigma}^{-1})^{-1}m\bm{\Sigma}^{-1} & \bm{\Upsilon}_{\bm{\theta\eta}}\bm{\Upsilon}_{\bm{\eta\eta}}^{-1}(n\bm{\Upsilon}_{\bm{\eta\eta}}^{-1}+m\bm{\Sigma}^{-1})^{-1}m\bm{\Sigma}^{-1} \\
\bm{0} & (n\bm{\Upsilon}_{\bm{\eta\eta}}^{-1}+m\bm{\Sigma}^{-1})^{-1}n\bm{\Upsilon}_{\bm{\eta\eta}}^{-1} & (n\bm{\Upsilon}_{\bm{\eta\eta}}^{-1}+m\bm{\Sigma}^{-1})^{-1}m\bm{\Sigma}^{-1}
\end{pmatrix}
\begin{pmatrix}
\tilde{\bm{\theta}} \\
\tilde{\bm{\eta}}   \\
\widehat{\bm{\eta}}
\end{pmatrix} + o_{p}(1).
\end{equation*}
Clearly, the fact that $n/(n+m)\rightarrow \gamma$ implies that $(n\bm{\Upsilon}_{\bm{\eta\eta}}^{-1}+m\bm{\Sigma}^{-1})^{-1}n\bm{\Upsilon}_{\bm{\eta\eta}}^{-1} \rightarrow \bm{W}_1 $ and $(n\bm{\Upsilon}_{\bm{\eta\eta}}^{-1}+m\bm{\Sigma}^{-1})^{-1}m\bm{\Sigma}^{-1} \rightarrow \bm{W}_2$ so we may rewrite the display above as 
\begin{equation} \label{Eq.Pf.II.No1}
\begin{pmatrix}
\bar{\bm{\theta}}_{C} \\
\bar{\bm{\eta}}_{C}
\end{pmatrix} = \bm{M}
\begin{pmatrix}
\tilde{\bm{\theta}} \\
\tilde{\bm{\eta}}   \\
\widehat{\bm{\eta}}
\end{pmatrix} +o_{p}(1)
\end{equation}
where $\bm{M}$ is given in \eqref{Eq.V&M}. Further observe that 
\begin{equation} \label{Eq.Pf.II.No2}
\bm{M}
\begin{pmatrix}
\bm{\theta}_{0} \\
\bm{\eta}_{0}  \\
\bm{\eta}_{0}
\end{pmatrix} = 
\begin{pmatrix}
\bm{\theta}_{0} \\
\bm{\eta}_{0}
\end{pmatrix}
\end{equation}
and that
\begin{equation} \label{Eq.Pf.II.No3}
\sqrt{n}(\tilde{\bm{\theta}}-\bm{\theta}_{0},\tilde{\bm{\eta}}-\bm{\eta}_{0},\widehat{\bm{\eta}}-\bm{\eta}_{0}) \Rightarrow \mathcal{N}_{p+2q}(\bm{0},\bm{V})     
\end{equation}
where $\bm{V}$ is given by \eqref{Eq.V&M}. Now \eqref{Eq.Pf.II.No1}, \eqref{Eq.Pf.II.No2} and \eqref{Eq.Pf.II.No3} together imply that
\begin{equation}
\begin{pmatrix}
\bar{\bm{\theta}}_{C} - \bm{\theta}_{0} \\
\bar{\bm{\eta}}_{C}  - \bm{\eta}_{0}
\end{pmatrix}
\Rightarrow \mathcal{N}_{p+q}(\bm{0},\bm{MVM}^T)
\end{equation}
as stated. In particular $\bm{C}_{\bm{\theta\theta}}$ is the appropriate submatrix of $\bm{MVM}^T$. Multiplying out we find that
\begin{equation} \label{Eq.4.Pf.4}
\bm{C}_{\bm{\theta\theta}} = \bm{\Upsilon}_{\bm{\theta\theta}} - \bm{\Upsilon}_{\bm{\theta\eta}}\bm{W}_{2}\bm{R}^{T}-\bm{R}\bm{W}_{2}\bm{\Upsilon}_{\bm{\eta\theta}}+ \bm{R}\bm{W}_{2}\bm{\Upsilon}_{\bm{\eta\eta}}\bm{W}_{2}\bm{R}^{T}+\rho \bm{R}\bm{W}_{2}\bm{\Sigma}\bm{W}_{2}\bm{R}^{T}.
\end{equation}
The matrices $\bm{\Upsilon}_{\bm{\eta\eta}}$, $\bm{\Sigma}$ and $\bm{W}_2$ are symmetric and thus their products commute. It follows that $\bm{R}\bm{W}_{2}\bm{\Upsilon}_{\bm{\eta\eta}}\bm{W}_{2}\bm{R}^{T}$ equals $\bm{R}\bm{\Upsilon}_{\bm{\eta\eta}}\bm{W}_{2}^{2}\bm{R}^{T}$ and $\rho \bm{R}\bm{W}_{2}\bm{\Sigma}\bm{W}_{2}\bm{R}^{T}$ equals $\rho \bm{R}\bm{\Sigma}\bm{W}_{2}^{2}\bm{R}^{T}$. It is also easy to verify that $\bm{\Upsilon}_{\bm{\theta\eta}}\bm{W}_{2}\bm{R}^{T}=\bm{R}\bm{W}_{2}\bm{\Upsilon}_{\bm{\eta\theta}}$ so 
$$
\bm{\Upsilon}_{\bm{\theta\eta}}\bm{W}_{2}\bm{R}^{T}+\bm{R}\bm{W}_{2}\bm{\Upsilon}_{\bm{\eta\theta}}=2\bm{\Upsilon}_{\bm{\theta\eta}}\bm{W}_{2}\bm{R}^{T}=2\bm{R}\bm{\Upsilon}_{\bm{\eta\eta}}\bm{W}_{2}\bm{R}^{T}.
$$
Combining and simplifying we obtain  
\begin{equation*} 
\bm{C}_{\bm{\theta\theta}} = \bm{\Upsilon}_{\bm{\theta\theta}} - \bm{R}\bm{\Upsilon}_{\bm{\eta\eta}}\bm{S}\bm{R}^{T}.
\end{equation*}
where
\begin{equation}
\bm{S}=2\bm{W}_{2}-\bm{W}_{2}^{2}-\rho\bm{W}_{2}^{2}\bm{\Sigma}\bm{\Upsilon}_{\bm{\eta\eta}}^{-1}.
\end{equation}
Now, using symmetry, standard algebraic manipulation and the fact that $\rho=\gamma/(1-\gamma)$ we have
\begin{align*}
\bm{S} &= (\gamma\bm{\Upsilon}_{\bm{\eta\eta}}^{-1}+(1-\gamma)\bm{\Sigma})^{-2} \{2(1-\gamma)\bm{\Sigma}^{-1}(\gamma\bm{\Upsilon}_{\bm{\eta\eta}}^{-1}+(1-\gamma)\bm{\Sigma})-(1-\gamma)^{2}\bm{\Sigma}^{-2}-\frac{\gamma}{1-\gamma}\bm{\Sigma}\bm{\Upsilon}_{\bm{\eta\eta}}^{-1}(1-\gamma)^{2}\bm{\Sigma}^{-2}\} \\
      &=(\gamma\bm{\Upsilon}_{\bm{\eta\eta}}^{-1}+(1-\gamma)\bm{\Sigma})^{-2}(1-\gamma)(\gamma\bm{\Upsilon}_{\bm{\eta\eta}}^{-1}+(1-\gamma)\bm{\Sigma}) \\
      &=(\gamma\bm{\Upsilon}_{\bm{\eta\eta}}^{-1}+(1-\gamma)\bm{\Sigma})^{-1}(1-\gamma)\bm{\Sigma}^{-1} = \bm{W}_{2}
\end{align*}
Thus $\bm{C}_{\bm{\theta\theta}} = \bm{\Upsilon}_{\bm{\theta\theta}} - \bm{R}\bm{\Upsilon}_{\bm{\eta\eta}}\bm{W}_{2}\bm{R}^{T} = \bm{\Upsilon}_{\bm{\theta\theta}} - \bm{\Upsilon}_{\bm{\theta\eta}}\bm{\Upsilon}_{\bm{\eta\eta}}^{-1}\bm{W}_{2}\bm{\Upsilon}_{\bm{\theta\eta}}^{T}$ as required. It is also clear that $\bm{C}_{\bm{\eta\eta}}=\bm{B}_{\bm{\eta\eta}}$ and that
\begin{align*}
\bm{C}_{\bm{\theta\eta}} &= \lim_{n}{\rm Cov}(\sqrt{n}(\bar{\bm{\theta}}_{C}-\bm{\theta}_{0}),\sqrt{n}(\bar{\bm{\eta}}_{C}-\bm{\eta}_{0})) = \lim_{n} n{\rm Cov}(\bar{\bm{\theta}}_{C},\bar{\bm{\eta}}_{C}) = \lim_{n} n{\rm Cov}(\tilde{\bm{\theta}}-\tilde{\bm{R}}(\tilde{\bm{\eta}}-\bar{\bm{\eta}}),\bar{\bm{\eta}})  \\
&= \lim_{n} n({\rm Cov}(\tilde{\bm{\theta}},\bar{\bm{\eta}})-\tilde{\bm{R}}{\rm Cov}(\tilde{\bm{\eta}}-\bar{\bm{\eta}},\bar{\bm{\eta}})) 
\end{align*}
where $\tilde{\bm{R}}=\tilde{\bm{\Upsilon}}_{\bm{\theta\eta}}\tilde{\bm{\Upsilon}}_{\bm{\eta\eta}}^{-1}$. Now $n{\rm Cov}(\tilde{\bm{\theta}},\bar{\bm{\eta}}) = n{\rm Cov}(\tilde{\bm{\theta}},\bm{W}_{1}\tilde{\bm{\eta}}+\bm{W}_2\widehat{\bm{\eta}}+o_{p}(1)) \rightarrow \bm{\Upsilon}_{\bm{\theta\eta}}\bm{W}_1$. Furthernote that 
$n{\rm Cov}(\tilde{\bm{\eta}}-\bar{\bm{\eta}},\bar{\bm{\eta}}) = n{\rm Cov}(\tilde{\bm{\eta}},\bm{W}_1\tilde{\bm{\eta}}+\bm{W}_2\widehat{\bm{\eta}}+o_{p}(1))-n{\rm Cov}(\bar{\bm{\eta}},\bar{\bm{\eta}}) \rightarrow \bm{\Upsilon}_{\bm{\eta\eta}}\bm{W}_1 - (\bm{\Upsilon}_{\bm{\eta\eta}}^{-1}+(\rho\bm{\Sigma})^{-1})^{-1}=\bm{0}$ since 
$$
\bm{\Upsilon}_{\bm{\eta\eta}}\bm{W}_1 =  \bm{\Upsilon}_{\bm{\eta\eta}}(\gamma\bm{\Upsilon}_{\bm{\eta\eta}}^{-1}+(1-\gamma)\bm{\Sigma}^{-1})^{-1}\gamma\bm{\Upsilon}_{\bm{\eta\eta}}^{-1}= (\bm{\Upsilon}_{\bm{\eta\eta}}^{-1}+(\rho\bm{\Sigma})^{-1})^{-1}
$$
where we have used the fact that $\rho = \gamma/(1-\gamma)$. Thus $\bm{C}_{\bm{\theta\eta}}=\bm{\Upsilon}_{\bm{\theta\eta}}\bm{W}_1$ concluding the proof. 
\end{proof}

\section*{{Proof of Theorem \ref{Thm.Order.C&U}:}}

\begin{proof}
Suppose that $(\bm{U},\bm{V})^{T}\sim\mathcal{N}((\bm{\theta},\bm{\eta})^{T},\bm{\Upsilon})$ and $\bm{W}\sim\mathcal{N}(\bm{\eta},\rho\bm{\Sigma})$ are independent. Let $I_{\bm{\omega}}(\bm{U},\bm{V})$ and $I_{\bm{\omega}}(\bm{U},\bm{V},\bm{W})$ denote the Fisher Information about $\bm{\omega}=(\bm{\theta},\bm{\eta})^{T}$ in $(\bm{U},\bm{V})$ and $(\bm{U},\bm{V},\bm{W})$ respectively. It is clear that $I_{\bm{\omega}}(\bm{U},\bm{V}) = \bm{\Upsilon}^{-1}$. Moreover, repeating the calculations in proofs of Lemma \ref{Lemma.Estimators.II} and Theorem \ref{Thm.Limit.4.Estimator.IIb} we deduce that $I_{\bm{\omega}}(\bm{U},\bm{V},\bm{W})=\bm{C}^{-1}$. The additivity of Fisher's Information implies that
\begin{equation} \label{Eq.FI}
I_{\bm{\omega}}(\bm{U},\bm{V},\bm{W}) \succeq I_{\bm{\omega}}(\bm{U},\bm{V}). 
\end{equation}
Equation \eqref{Eq.FI} and Lemma \ref{Lemma.Loewener} imply that
\begin{equation*}
\bm{C} \preceq \bm{\Upsilon}    
\end{equation*}
as stated. The fact that $\bm{V}_{\bm{C}}^{\bm{\Phi}} \preceq \bm{V}_{\bm{\Upsilon}}^{\bm{\Phi}}$ now follows as in Theorem \ref{Thm.Order.A&B}. 
\end{proof}

\section*{{Proof of Theorem \ref{Thm.Order.B&C}:}}

\begin{proof}
By Equation \eqref{Eq.II.Pf.II} in the proof of Theorem \ref{Thm.Limit.4.Estimator.IIa} we have
\begin{equation*} 
\sqrt{n}(\bar{\bm{\theta}}_{B}-\bm{\theta }_{0})=-
(\bm{D}_{\bm{\theta}_{0}}(\bm{\psi}))^{-1}\{\frac{1}{\sqrt{n}}\sum_{i=1}^{n}\bm{\psi}(\bm{\theta }_{0},\bm{\eta}_{0},\bm{Y}_{i})+ \bm{D}_{\bm{\eta}_0}(\bm{\psi})\sqrt{n}(\bar{\bm{
\eta }}-\bm{\eta }_{0})\}+o_{p}\left( 1\right),
\end{equation*}
and similarly, 
\begin{equation*} 
\sqrt{n}(\bar{\bm{\theta}}_{C}-\bm{\theta }_{0})=-
(\bm{D}_{\bm{\theta}_{0}}(\bm{\lambda}))^{-1}\{\frac{1}{\sqrt{n}}\sum_{i=1}^{n}\bm{\lambda}(\bm{\theta }_{0},\bm{\eta}_{0},\bm{Y}_{i})+ \bm{D}_{\bm{\eta}_0}(\bm{\lambda})\sqrt{n}(\bar{\bm{
\eta }}-\bm{\eta }_{0})\}+o_{p}\left( 1\right).
\end{equation*}
The analysis in the proof of Theorem \ref{Thm.Limit.4.Estimator.IIa} shows that in both equations above the terms in the curly brackets are asymptotically independent. Conditions \eqref{Eq.Efficient} and \eqref{Eq.Sensitive} immediately imply the conclusion of the Theorem.  
\end{proof}

By Equation \eqref{Eq.II.Pf.II} in the proof of Theorem \ref{Thm.Limit.4.Estimator.IIa} 

\begin{remark} \label{Rem.More.B&C}
Recall that $(\tilde{\bm{\theta}},\tilde{\bm{\eta}})$ simultaneously solve $\bm{\Psi}(\bm{\theta},\bm{\eta})=\bm{0}$ and $\bm{\Gamma}(\bm{\theta},\bm{\eta})=\bm{0}$ where $\bm{\Gamma}(\bm{\theta},\bm{\eta})=n^{-1} \sum_{i=1}^{n}\bm{\gamma}(\bm{\theta },\bm{\eta},\bm{Y}_{i})$. Standard calculations show that
\begin{equation} \label{Eq.I.More.B&C}
\begin{pmatrix}
\tilde{\bm{\theta}} \\
\tilde{\bm{\eta}}
\end{pmatrix} =     
\begin{pmatrix}
\bm{\theta}_{0} \\
\bm{\eta}_{0}
\end{pmatrix} + \bm{D}^{-1}
\begin{pmatrix}
\frac{1}{n}\sum_{i=1}^{n}\bm{\psi}(\bm{\theta },\bm{\eta},\bm{Y}_{i}) \\
\frac{1}{n}\sum_{i=1}^{n}\bm{\gamma}(\bm{\theta },\bm{\eta},\bm{Y}_{i})
\end{pmatrix} + o_{p}(1)
\end{equation}
where
\begin{equation*}
\bm{D} =
\begin{pmatrix}
\bm{D}_{11} & \bm{D}_{12} \\
\bm{D}_{21} & \bm{D}_{22}
\end{pmatrix} =
\begin{pmatrix}
\bm{D}_{\bm{\theta}}(\bm{\psi}) & \bm{D}_{\bm{\eta}}(\bm{\psi}) \\
\bm{D}_{\bm{\theta}}(\bm{\gamma}) & \bm{D}_{\bm{\eta}}(\bm{\gamma})
\end{pmatrix}.
\end{equation*}
Using the above notations and rewriting Equation \eqref{Eq.II.Pf.II} we have
\begin{equation}
\bar{\bm{\theta}}_{B} = \bm{\theta}_{0} + \frac{1}{n}\sum_{i=1}^{n} \bm{D}_{11}^{-1}\bm{\psi}(\bm{\theta },\bm{\eta},\bm{Y}_{i})-\bm{D}_{11}^{-1}\bm{D}_{12}(\bar{\bm{\eta}}-\bm{\eta}_0)+o_{p}(1).   
\end{equation}
As demonstrated in the proof of Theorem \ref{Thm.Limit.4.Estimator.IIa} the two terms above are asymptotically independent so we can re-express $\bar{\bm{\theta}}_{B}$ as 
\begin{equation} \label{Eq.II.More.B&C}
\bar{\bm{\theta}}_{B} = \bm{\theta}_{0} + \frac{1}{n}\sum_{i=1}^{n} 
\{\bm{D}_{11}^{-1}\bm{\psi}(\bm{\theta },\bm{\eta},\bm{Y}_{i})-\bm{D}_{11}^{-1}\bm{D}_{12}\bm{Q}_i\}+o_{p}(1)  
\end{equation}
where $\bm{Q}_i$ are IID $\mathcal{N}(\bm{0},\bm{W}_1\bm{\Upsilon}_{\bm{\eta\eta}} \bm{W}_1^{T} + \rho \bm{W}_2\bm{\Sigma}\bm{W}_2^{T})$ RVs which are independent of $\mathcal{D}$. Furthernote that by \eqref{Eq.I.More.B&C}
\begin{align*}
\tilde{\bm{\theta}} &= \bm{\theta}_{0} +\bm{D}^{11}\frac{1}{n}\sum_{i=1}^{n}\bm{\psi}(\bm{\theta },\bm{\eta},\bm{Y}_{i})+\bm{D}^{12}\frac{1}{n}\sum_{i=1}^{n}\bm{\gamma}(\bm{\theta },\bm{\eta},\bm{Y}_{i})+o_p(1),\\
\tilde{\bm{\eta}} &= \bm{\eta}_{0} +\bm{D}^{21}\frac{1}{n}\sum_{i=1}^{n}\bm{\psi}(\bm{\theta },\bm{\eta},\bm{Y}_{i})+\bm{D}^{22}\frac{1}{n}\sum_{i=1}^{n}\bm{\gamma}(\bm{\theta },\bm{\eta},\bm{Y}_{i})+o_p(1),\\
\end{align*}
where $\bm{D}^{ij}$ is the appropriate submatrix of $\bm{D}^{-1}$. Substituting the formulas above into Equation \eqref{Eq.theta.bar} for $\bar{\bm{\theta}}_{C}$ and simplifying we find that
\begin{equation} \label{Eq.III.More.B&C}
\bar{\bm{\theta}}_{C} = \bm{\theta}_{0} + \frac{1}{n}\sum_{i=1}^{n} 
\{
(\bm{D}^{11}-\bm{\Upsilon_{\theta\eta}}\bm{\Upsilon}_{\bm{\eta\eta}}^{-1}\bm{D}^{21})\bm{\psi}(\bm{\theta },\bm{\eta},\bm{Y}_{i})+(\bm{D}^{21}-\bm{\Upsilon_{\theta\eta}}\bm{\Upsilon}_{\bm{\eta\eta}}^{-1}\bm{D}^{22})\bm{\gamma}(\bm{\theta },\bm{\eta},\bm{Y}_{i})+\bm{\Upsilon_{\theta\eta}}\bm{\Upsilon}_{\bm{\eta\eta}}^{-1}\bm{Q}_i\}+o_p(1).
\end{equation}
Therefore comparing the estimators $\bar{\bm{\theta}}_{B}$ and $\bar{\bm{\theta}}_{C}$ amounts to comparing their influence functions implicit in \eqref{Eq.II.More.B&C} and \eqref{Eq.III.More.B&C}, i.e.,
$$
\bm{D}_{11}^{-1}\bm{\psi}(\bm{\theta },\bm{\eta},\bm{Y}_{i})-\bm{D}_{11}^{-1}\bm{D}_{12}\bm{Q}_i
$$
and 
$$
(\bm{D}^{11}-\bm{\Upsilon_{\theta\eta}}\bm{\Upsilon}_{\bm{\eta\eta}}^{-1}\bm{D}^{21})\bm{\psi}(\bm{\theta },\bm{\eta},\bm{Y}_{i})+(\bm{D}^{21}-\bm{\Upsilon_{\theta\eta}}\bm{\Upsilon}_{\bm{\eta\eta}}^{-1}\bm{D}^{22})\bm{\gamma}(\bm{\theta },\bm{\eta},\bm{Y}_{i})+\bm{\Upsilon_{\theta\eta}}\bm{\Upsilon}_{\bm{\eta\eta}}^{-1}\bm{Q}_i,
$$
respectively. Also note that 
\begin{equation*}
\bm{\Upsilon}=(\bm{D}^{-1})
\begin{pmatrix}
\mathbb{E}_{0}(\bm{\psi\psi}^T) & \mathbb{E}_{0}(\bm{\psi\gamma}^T) & \\ \mathbb{E}_{0}(\bm{\gamma\psi}^T) & \mathbb{E}_{0}(\bm{\gamma\gamma}^T)
\end{pmatrix}
(\bm{D}^{-1})^{T}    
\end{equation*}
so although in principal it is possible to always compare the above influence functions in practice this comparison is very difficult unless some further simplifying assumptions are imposed. 
\end{remark}

\end{document}